\documentclass[english,tightenlines,eqsecnum,amsmath,amssymb,aps,prd,superscriptaddress,nofootinbib,showpacs,floats]{article}
\usepackage{amsmath,amsfonts,amssymb,multicol}
\usepackage{graphicx,caption,subcaption}
\usepackage[usenames]{xcolor}
\definecolor{AHZ}{rgb}{0.0,0.3,0.2}
\usepackage[colorlinks,linkcolor=AHZ,citecolor=red,bookmarks]{hyperref}
\usepackage{rotating}
\usepackage[mathscr]{eucal}
\usepackage{graphicx}
\usepackage[T1]{fontenc}
\usepackage{babel}
\usepackage{authblk}
\usepackage{epsfig}
\usepackage{caption, threeparttable}
\usepackage{amsmath,amsfonts,amssymb,multicol}
\usepackage{graphicx,caption,subcaption}
\usepackage[usenames]{xcolor}
\definecolor{AHZ}{rgb}{0.0,0.3,0.2}
\usepackage[colorlinks,linkcolor=AHZ,citecolor=red,bookmarks]{hyperref}
\usepackage{rotating}
\usepackage[mathscr]{eucal}
\usepackage{graphicx}
\usepackage[T1]{fontenc}
\usepackage{babel}
\usepackage{authblk}
\usepackage{epsfig}
\usepackage{color}
\usepackage{amssymb}
\usepackage{fancyhdr}

\begin{document}

\title{Cosmological consequences and statefinder diagnosis of a non-interacting generalized Chaplygin gas in $f(R,T)$ gravity}
\author{Hamid Shabani,\thanks{h.shabani@phys.usb.ac.ir}}
\affil{Department of Physics, University of Sistan and Baluchestan, Zahedan, Iran}
\date{\today}
%\preprint{hep-th/yymmnnn}
%
\maketitle
\begin{abstract}
\noindent
In this paper, we investigate cosmological consequences of a scenario
for recently reported accelerated expansion of the Universe, in which the
generalized Chaplygin gas (GCG) along with the baryonic matter are
responsible for this observed phenomenon. In the present model, we employ
an isotropic and homogeneous FLRW space time in the framework of $f(R,T)$
theory of gravity. In $f(R,T)$ gravity, the conservation of energy-momentum
tensor (EMT) leads to a constraint equation which enforces us to use some
specific forms of type $f(R,T)=g(R)+h(T)$. In this work we choose $g(R)=R$.
We consider three classes of Chaplygin gas models which include three
different forms of $f(R,T)$ function; those models which employ the
standard Chaplygin gas (SCG), models which use GCG in the high pressure
regimes and finally, the third case is devoted to investigating high density
regimes in the presence of GCG. The effective equation of state (EoS) and the
deceleration parameter for these Chaplygin gas models are calculated and it is
indicated that the related present values can be observationally acceptable in
$f(R,T)$ gravity. Among the models we shall present, type two models has a
better situation; the predictions of these models are more consistent with
the observational data. We also examine these models via the statefinder
diagnostic tool. The statefinder parameters $s$ and $r$ are obtained for
these three scenarios and various trajectories are plotted for different
values of the model parameters. The involved parameters are, $K$ and $\alpha$
which appear in the GCG density and the parameters $\mathbf{m}$ and $\mathbf{n}$
which are related to the integral constants which are appeared
in the process of obtaining $h(T)$. The discrimination between different dark energy
models is considered by varying the mentioned parameters for these three
types of models. Since positive values for $\mathbf{m}$ and
$\mathbf{n}$'s would lead to some divergences, we will take negative
values. It is discussed that the consistent values for the effective
EoS for all allowed values of $K$, can only be obtained in the high
pressure regimes. We show that the distance of these models (which is defined as
the difference between the predicted present values of the statefinder
parameters of a model and the corresponding ones for $\Lambda$CDM model),
varies for different choices of the model parameters. However,
the distance does not depend on the values of $\alpha$ for models of type three.
Finally, we test these Chaplygin gas models using recent Hubble parameter
as well as type Ia supernova data. We find that these models
are compatible with type Ia supernova data. Particularly, we plot
the modulus distance, the Hubble parameter and the effective EoS
parameter for the best fit values of the model(s) parameters. Also,
we compare the predicted present values of the statefinder parameters
to the observational data.
\end{abstract}
%\pacs{98.80.-k; 04.50.Kd; 95.36.+x; 98.80.Jk}
%\keywords{Cosmology; $f(R,T)$ Gravity; Dark Energy; Chaplygin gas; statefinder diagnosis.}
%
\section{Introduction}\label{intro}
Currently, observational experiments show that the Universe is undergoing an
accelerated expansion~\cite{supno1,supno2,supno3,WMAP1,Teg1,SDSS1,SDSS2,WMAP2,WMAP3,WMAP4,WMAP5}.
The most accepted agreement is that there is two components which play a serious and
important role in the formation and evolution of the Universe; the so called ``dark
matter" (DM), which is responsible for the structure formation and clustering of the
galaxies~\cite{dmatt1,dmatt2,dmatt3,dmatt4,dmatt5,dmatt6,dmatt7,dmatt8} and the ``dark
energy" (DE) which makes a negative pressure and thus gives rise to the accelerated
expansion of the Universe~\cite{dener1,dener2,dener3,dener4,dener5,dener6,dener7,
dener8,dener9,dener10,dener11,dener12}. The former has a contribution of about
$26$\% of the total matter density, the latter forms about $69$\% of it and the rest
is related to the baryonic (visible) matter~\cite{Planck1,Planck2}.
There are two general approaches which can be used to
face the problems associated to DE and DM. The first one deals with introducing some
matter field(s) that affect the dynamical evolution of the universe and thereby these
observed phenomena may be explained. In the second approach one can alter the geometrical
sector of GR to obtain some theoretical results corresponding to the observational ones.
In order to apply the former, numerous scenarios have been proposed so far. The cosmological
constant can be accounted for as the most successful and important candidate for DE which has
a constant EoS parameter. The GR theory comprised by the cosmological constant and DM is
called the concordance or the $\Lambda \rm{CDM}$ model~\cite{LCDM}. In spite of high consistency
of $\Lambda \rm{CDM}$ model with the present observations, it suffers from an important problem;
the theoretical value of the DE density is not compatible with the observationally accepted
one. Indeed, their values differ about $120$ orders of magnitude. This problem is called
``the cosmological constant problem"~\cite{Cos.pro1,Cos.pro2,Cos.pro3,Cos.pro4,Cos.pro5,
Cos.pro6,Cos.pro7,Cos.pro8}. The other proposals for the accelerated expansion of the
Universe are the dynamical DE models with time-varying EoS e.g., quintessence and phantom
models~\cite{quint1,quint2,quint3,quint4,quint5,quint6}. However, these models may also
include other problems, e.g., the ``fine tuning" which is a major issue, by which the
quintessence models are afflicted. Another candidate for DE (which is the main subject of
our study) is the Chaplygin gas (CG) model introduced as to explain
the DE phenomenon~\cite{CG1,CG2}. Following a perfect fluid, the EoS for this matter
is, $p=-K /\rho$ where $K$ is a constant (this case is called the standard Chaplygin
gas (SCG)). It has been shown that the Chaplygin cosmology can be interpreted as a
transition from a DM dominated universe to an accelerated expansion universe including
an intermediate phase with ``stiff fluid" (which corresponds to a fluid with $p=\rho$)
domination~\cite{CG3}. Although, the SCG model can lead to a transition from a decelerated
to accelerated expansion, it cannot explain the structure formation and it is encountered
some problems with the cosmological power spectrum~\cite{CGprob1,CGprob2}. Consequently,
the EoS of the SCG model has been generalized as $p=-K /\rho^{-\alpha}$, to confront the
mentioned issues. Note that, the general CG model (GCG) gives rise to an intermediate
epoch with the EoS $p=\alpha\rho$~\cite{CG2}. The GCG model has been used in different
areas such as, the DE problem~\cite{GCG1,GCG2}, Matter power spectrum~\cite{GCG3} and
wormholes~\cite{GCG4,GCG5,GCG6}.

The geometrical manipulation of GR is the other approach with the aim of solving some problems
such as DE and DM. In the past decades, higher order gravity, specially $f(R)$ gravity has been
introduced by replacing the Ricci scalar in the Einstein-Hilbert action with some scalar
invariants. This model has drawn a remarkable attention among the other models and theories.
In $f(R)$ theory of gravity, one simply replaces the Ricci curvature scalar with a function of it. This
theory has had successes and failures in different aspects~\cite{fR1,fR2,fR3,fR4,fR5,fR6,fR7,
fR8,fR9,fR10}. As an idea toward the development of $f(R)$ gravity, some authors assume that
the matter Lagrangian is non-minimally coupled to the geometrical sector of the
action~\cite{frL1,frL2,frL3,frL4,frL5}. In Ref.~\cite{frL6} a more complete form of
these type of models entitled as $f(R,L_{m})$ gravity is proposed. Another suggestion
to couple the matter sector to the geometrical one is to employ the trace of EMT. This
model was presented by authors of Ref.~\cite{fRT1}, and various features of
this model have been reported in~\cite{fRT2,fRT3,fRT4,fRT5,fRT6,fRT7,fRT8,fRT9,fRT10,
fRT11,fRT12,fRT13,fRT14,fRT15}.

In the present work we use both approach, namely, we work on a model in which GCG and the baryonic matter
are considered as the whole matter in $f(R,T)$ gravity as the background geometric law. Simultaneously
use of both approach have been already employed in the literature to describe the early or late time
phenomena in cosmology~\cite{elmardi,borowiec}. Since $f(R,T)$ gravity introduces a new approach to
involve the interactions of matter effects with spacetime curvature, it is well-motivated to consider
cosmological consequences of different types of matter in the framework of this theory. The GCG in the
background of GR involved some issues e.g., the standard case, namely SCG has ruled out
by the most observational data or in some works it is reported that only specific subclasses of solutions
would pass the observational tests (i.e., those with $\alpha\sim0$)~\cite{rul1,rul2,rul3,rul4,rul5,rul6}.
Authors of Ref.~\cite{sandvik} have shown that CG produces oscillations or exponential blowup of the
power spectrum of dark matter which is not consistent with observations. Particularly, they
have reported that $99.999\%$ of GCG parameter space which previously has been allowed, would
be excluded. However, in Ref.~\cite{popolo} it has been discussed that considering the
joint effect of shear and rotation can overcome this instability. In this regard, the study of
Chaplygin gas in other frameworks such as $f(R,T)$ gravity, in search of a viable dark
energy model may leads to more desirable results.
Cosmological considerations of every theory of gravity are based on the investigation
of the behavior of the Hubble and the deceleration parameters which are defined as the
first and second order of time derivatives of the scale factor i.e., $H=\dot{a}/a$,
$q=\ddot{a}/(aH^{2})$. The former determines the expansion of the Universe and the latter
indicates the acceleration/deceleration behavior. The DE models are constructed so as to
match them with the observations. For this reason, the most consistent DE models give the
same present values of $H$ and $q$. Therefore, there is some type of degeneracy in the
present values, $H_{0}$ and $q_{0}$. To deal with this issue, various strategies have been
proposed. The statefinder diagnosis is an efficient tool to discriminate different DE
models. The authors of Ref.~\cite{stfi1} introduced new cosmological parameters
constructed out of the third order of time derivative of the scale factor which are
called ``statefinder parameters" and are denoted by the pair
($s\equiv(r-1)/3(q-1/2),r\equiv\ddot{a}/(aH^{3})$). Since, this tool use the different
orders of the time derivative of the scale factor, it is a geometrical diagnosis. This
method can distinguish between DE models that have the same values of $H_{0}$ and $q_{0}$,
in a higher level. Note that this method can be extended to the models with higher degrees
of degeneracy. In the statefinder diagnostic tool, the DE models indicated by the ($s,r$)
plane trajectories and in this way one could consider the behavior of different DE models
and thus discriminate them. We have mentioned that the $\Lambda\rm{CDM}$ model is still good
fitted to the observation and therefore the suggested DE models should not be so far from
this model. To understand this fact, the difference between the predicted present values
$s_{0}$ and $r_{0}$ of models and the corresponding values of $\Lambda\rm{CDM}$ model
($s_{0}^{(\Lambda\rm{CDM})},r_{0}^{(\Lambda\rm{CDM})}$) is a criteria for distinction of the
DE models (this difference is called ``distance of model").

Up to now, different DE models are diagnosed which are mainly of scalar field type models, such
as $\Lambda\rm{CDM}$ and quintessence DE models which have been considered in~\cite{stfi1,stfi2}
and are accounted for as the first research works that introduced the statefinder diagnostic tool,
the interacting quintessence models~\cite{stfi3,stfi4}, the holographic
DE models~\cite{stfi5,stfi6}, the holographic DE model in a non-flat universe~\cite{stfi7},
the phantom model~\cite{stfi8}, the tachyon model~\cite{stfi9}, the GCG model~\cite{stfi10} and
the interacting new agegraphic DE model in a flat and a non-flat universe~\cite{stfi11,stfi12}.

We plan our investigations as follows: in Sec.~\ref{Fieldequations}, we start with presenting
the action of $f(R,T)$ gravity and obtain the related field equations and a constraint that
must be satisfied by the function $f(R,T)$ in order to guarantee the conservation of EMT. We
devote the rest of this section to present the procedure of obtaining the function $f(R,T)$.
For some reasons we presume a linear combination of the matter and curvature sectors of
the Lagrangian, i.e., $f(R,T)=R+h(T)$. This choice facilitates the procedure of obtaining
$f(R,T)$ function using the EMT conservation and makes us capable of getting exact solutions.
Minimal couplings respect the EMT conservation and for these type of models all equations and
solutions can be easily reduced to GR. These models can be seen as a modification to GR. Also,
the models including minimal couplings respect the equivalence principle, in spite of non-minimal
ones which show some violation, in this respect~\cite{haghani,Sharif}.
We get three classes of $f(R,T)$ function. Class I for the SCG model, class II for the GCG
model in high pressure situations and class III for the GCG model in high density
regimes. In Sec.~\ref{eq-def} we introduce the statefinder parameters and obtain them for these
three models. Moreover, we bring forward some cosmological parameters that are helpful for the
rest of our study. In Sec.~\ref{numerical}, we
present adequate numerical plots to show and discuss the cosmological consequences of these
three models. Sec.~\ref{test} has been devoted to testing models I, II and III using the recent
observational data via the chi-square technique and finally, in Sec.~\ref{conclusions} we summerize
our results.
\section{Field equations of $f(R,T)$ gravity and the conservation of EMT}\label{Fieldequations}
In this section we present the field equation of $f(R,T)$ theory of gravity
and also discuss the conservation of EMT when two
pressureless baryonic matter and GCG\footnote{As we shall discus, GCG has a
dual feature in the past and future. It behaves like the cold dark matter (CDM) in
the early times and DE in the late times. Thus, we use
GCG as it plays the role of dark sector of the Universe.} are taken into account.
By including the two forms of matters, the action of $f(R,T)$ theory of gravity
can be written as
\begin{align}\label{action}
S=\int \sqrt{-g} d^{4} x \left[\frac{1}{16 \pi G} f\Big{(}R, T^{\textrm{(b, G)}}\Big{)}
+L^{\textrm{(total)}} \right],
\end{align}
where we have defined the Lagrangian of the total matter as
\begin{align}\label{lagrangian}
L^{\textrm{(total)}}\equiv L^{\textrm{(b)}}+L^{\textrm{(G)}}.
\end{align}
In the above equations $R$, $T^{\textrm{(b, G)}}\equiv g^{\mu \nu}
T^{\textrm{(b, G)}}_{\mu \nu}$, $L^{\textrm{(b, G)}}$ are the Ricci
curvature scalar, the trace of EMT of the baryonic matter and GCG (which we get
these fluids as the total matter content of the Universe) and the Lagrangian of the total matter,
respectively. The superscripts $b$ and $G$ stand for the baryonic matter and GCG,
$g$ is the determinant of the metric and we set $c=1$. The EMT $T_{\mu \nu}
^{\textrm{(b, G)}}$ is defined as the Euler-Lagrange expression of the
Lagrangian of the total matter, i.e.,
\begin{align}\label{Euler-Lagrange}
T_{\mu \nu}^{\textrm{(b, G)}}\equiv-\frac{2}{\sqrt{-g}}
\frac{\delta\left[\sqrt{-g}(L^{\textrm{(b)}}+L^{\textrm{(G)}})
\right]}{\delta g^{\mu \nu}}.
\end{align}

Instead of obtaining the field equations for the action (\ref{action}),
it is sufficient to obtain the corresponding field equations of
$f(R,T)$ gravity for general trace $T$ and the matter Lagrangian $L^{(m)}$,
and then solve them using the related quantities of the two forms of matters;
namely, the baryonic matter and GCG. If in the action (\ref{action}), the
trace $T^{\textrm{(b, G)}}$ and $L^{\textrm{(total)}}$ are replaced by trace
$T$ and $L^{\textrm{(m)}}$, respectively, the field equations for $f(R,T)$
gravity can be obtained as~\cite{fRT1}
\begin{align}\label{fRT field equations}
F(R,T) R_{\mu \nu}-\frac{1}{2} f(R,T) g_{\mu \nu}+\Big{(} g_{\mu \nu}
\square -\triangledown_{\mu} \triangledown_{\nu}\Big{)}F(R,T)=\Big{(}8
\pi G-{\mathcal F}(R,T)\Big{)}T_{\mu \nu}-\mathcal {F}(R,T)\mathbf
{\Theta_{\mu \nu}},
\end{align}
where
\begin{align}\label{theta}
\mathbf{\Theta_{\mu \nu}}\equiv g^{\alpha \beta}\frac{\delta
T_{\alpha \beta}}{\delta g^{\mu \nu}}~~~~~
~~~~~\mbox{and}~~~~~~~~~~
T_{\mu \nu}\equiv-\frac{2}{\sqrt{-g}}
\frac{\delta\Big{(}\sqrt{-g}L^{\textrm{(m)}}\Big{)}}{\delta g^{\mu \nu}},
\end{align}
and, for the sake of convenience, we have defined the following
functions for the derivatives with respect
to the trace $T$ and the Ricci curvature scalar $R$
\begin{align}\label{f definitions1}
{\mathcal F}(R,T) \equiv \frac{\partial f(R,T)}{\partial T}~~~~~
~~~~~\mbox{and}~~~~~~~~~~
F(R,T) \equiv \frac{\partial f(R,T)}{\partial R}.
\end{align}
Assuming a perfect fluid and a spatially flat
Friedmann--Lema\^{\i}tre--Robertson--Walker (FLRW) metric
\begin{align}\label{metricFRW}
ds^{2}=-dt^{2}+a^{2}(t) \Big{(}dx^{2}+dy^{2}+dz^{2}\Big{)},
\end{align}
together with using equation (\ref{fRT field equations}), the generalized
Friedmann equation for a perfect fluid with non-zero pressure is
achieved as
\begin{align}\label{Friedmann}
3H^{2}F(R,T)+\frac{1}{2} \Big{(}f(R,T)-F(R,T)R\Big{)}+3\dot{F}
(R,T)H=\Big{(}8 \pi G +{\mathcal F}
(R,T)\Big{)}\rho+\mathcal {F}(R,T)p.
\end{align}

Applying the Bianchi identity to the field equation
(\ref{fRT field equations}) leads to
the following constraint for multi perfect fluids
\begin{align}\label{constraint1}
\sum_{i=1}^{\mathcal{N}}\dot{\mathcal {F}_{i}}(R,T)(\rho_{i}+p_{i})-
\frac{1}{2}\mathcal {F}_{i}(R,T)(\dot{p_{i}}-\dot{\rho_{i}})=0,
\end{align}
where $\mathcal{N}$ is the number of included perfect fluids. Constraint
(\ref{constraint1}) for a pressureless matter reduces to\footnote{For
more details see Refs.~\cite{fRT7,fRT8}. }
\begin{align}\label{constraint2}
\dot{\mathcal {F}}(R,T)=\frac{3}{2}H(t)\mathcal {F}(R,T).
\end{align}
Equations (\ref{constraint1}) and (\ref{constraint2})
are two constraints that restrict the form of the Lagrangian
density $f(R,T^{\textrm{(b, G)}})$ in our case. These constraints
guarantee the conservation of EMT
which is imposed by the Bianchi
identity\footnote{Briefly, one can recast the field
equations in a canonical form similar to GR.
In this case, all the other terms can be
realized as the matter terms, e.g. see Refs.~\cite{fRT7,fRT8}.}.
Therefore, to obtain the Hubble parameter from equation
(\ref{Friedmann}), we need to find the function
$f(R, T^{\textrm{(b, G)}})$ that satisfy these constraints.

Generally, function $f(R, T^{\textrm{(b, G)}})$ can take an
arbitrary form in equations (\ref{Friedmann}) and (\ref{constraint1}).
As a matter of fact, viable models of $f(R,T)$ gravity can be classified
in the following manner~\cite{fRT1}
\begin{itemize}
  \item The $f(R,T)=R+h(T)$ models which have been mostly considered in the literature.
        We can enumerate some motivations to propose these type of models; (i) equations
        of motion and the related calculations are more tractable in the framework of
        these models as compared to the other ones, such that one can always simply
        obtain the GR results by switching off the function $h(T)$. From mathematical
        point of view, in each theory modest field equations can result in exact
        solutions without resorting to numerical methods. (ii) these models can
        be seen as a trivial modification to GR which depends on the matter
        content (regardless of the simplest possibility, i.e., adding a cosmological constant).
        In this regard, a simple correction may lead to some plausible results comparing with the other type
        of corrections. These models may also connect to other theories, for example, those in
        which the cosmological constant clearly depends on the trace of EMT which is called ``$\Lambda(T)$
        gravity" in the literature~\cite{LamdaT-1,LamdaT-2,LamdaT-3}. (iii) since any theory
        of modified gravity must be reduced to the Einstein-Hilbert action at low curvature
        regimes, for more complicated $f(R,T)$ functions, GR may not be recovered.
        (iv) these models are well-defined for cosmological fluids with $T=0$, like
        radiation (as we will see the third type of models may show some
        inconsistencies). (v) the EMT conservation tells us that for a linear combination
        of curvature and matter sectors we can determine the form of $h(T)$ function
        up to some integration constants. However, for complicated forms, this option
        may not be accessible. Particularly, in our work the EMT conservation leads to an
        intricate function for the standard Chaplygin gas, see expression (\ref{fRT-I}).
        Therefore a reasonable choice is to consider the contribution due to the trace
        of EMT into Lagrangian density, linearly. Such a setting makes calculations
        rough enough to take the Ricci scalar instead of an arbitrary function of
        it, as the geometrical part of the action.
  \item Models with arbitrary forms for the Ricci scalar but minimal coupling to
        the EMT trace: $f(R,T)=g(R)+h(T)$. These cases can be accounted for
        corrections on $f(R)$ gravity which may cure possible deficiencies
        of these theories. As some of the above motivations are still
        mentionable for these class of models, the complexities involved
        in the $g(R)$ function may lead to some problems such as finding exact
        solutions. In the present work, the benefit of choosing the simple
        form $g(R)=R$ is to obtain the Hubble parameter in terms of the
        matter terms, in a straightforward way. We note that the exact form of
        the Hubble parameter is needed in order to calculate the statefinder
        parameters, see e.g., equation (\ref{Hubble}). It is obvious that
        from equation (\ref{Friedmann}), for cases in which
        $F(R,T)\neq constant$ the Hubble parameter may not be obtained exactly.
  \item The models with $f(R,T)=g_{1}(R)+h(T)g_{2}(R)$. The field equations for
        these type of models can be so complicated due to non-minimal
        coupling term. These categories suffer from an important problem.
        Since for an ultra relativistic fluid the trace of EMT is vanished,
        the models with non-minimal coupling between the geometrical and matter
        sectors in the Lagrangian can become singular. Optimistically, these cases
        reduce to $f(R)$ gravity similar to the cases discussed in the above
        items. For instance, the actions with a pure non-minimal
        term (i.e., $g_{1}(R)=0$) can become either singular (eg., $h(T)\propto Log (T)$)
        or null, since the trace of ultra relativistic matter is vanished.
        Another issue is that, the excellent information about the function $h(T)$ which can
        be extracted from the models with linear combinations may be lost even for simple
        forms of $h(T)g_{2}(R)$. As a result, equation (\ref{constraint2}) turns into
        a completely complicated expression in terms of second time derivative of the Hubble
        parameter once non-minimal couplings are allowed. This in turn leaves us
        without any reasonable outcome for $h(T)$ function. The major concern of theories
        including non-minimal couplings in the Lagrangian is that they could lead to the
        violation of the equivalence principle~\cite{bertolami}.
\end{itemize}

Motivated by the above discussion, we
consider a function of the following form
\begin{align}\label{fRT function}
f\Big{(}R, T^{\textrm{(b, G)}}\Big{)}=R+h_{1}\Big{(}T^{\textrm{(b)}}\Big{)}+h_{2}
\Big{(}T^{\textrm{(G)}}\Big{)}.
\end{align}
Using definitions (\ref{f definitions1}) and the ansatz (\ref{fRT function}) we get
\begin{align}\label{f definitions2}
&\mathcal {F}_{1}\Big{(}R, T^{\textrm{(b, G)}}\Big{)}=
\frac{\partial f\Big{(}R, T^{\textrm{(b, G)}}\Big{)}}
{\partial T^{\textrm{(b)}}}=
h_{1}'\Big{(}T^{\textrm{(b)}}\Big{)},\nonumber\\
&\mathcal {F}_{2}\Big{(}R, T^{\textrm{(b, G)}}\Big{)}=
\frac{\partial f\Big{(}R, T^{\textrm{(b, G)}}\Big{)}}
{\partial T^{\textrm{(G)}}}=
h_{2}'\Big{(}T^{\textrm{(G)}}\Big{)},\nonumber\\
&F\Big{(}R, T^{\textrm{(b, G)}}\Big{)}=
\frac{\partial f\Big{(}R, T^{\textrm{(b, G)}}
\Big{)}}{\partial R}=1,
\end{align}
where the prime denotes a derivative with respect
to the argument. Applying the derivatives
(\ref{f definitions2}) to equation
(\ref{Friedmann}) leaves us with the following relation
\begin{align}\label{Hubble}
3H^{2}=\sum_{i=1}^{2} \Big{(} 8\pi G \rho_{i} +\mathcal
{F}_{i}(\rho_{i}+p_{i}) -\frac{1}{2}h_{i}\Big{)},
\end{align}
where the summation should be done over all terms corresponding
to the two forms of matter. Notice that,
hereafter the arguments in
$\mathcal {F}_{i}(R, T^{\textrm{(b, G)}})$,
$h_{1}(T^{\textrm{(G)}})$ and
$h_{2}(T^{\textrm{(b)}})$ will be
dropped for simplicity until we may restore them for some purposes.

For a perfect fluid with $p=p(\rho)$, the constraint
(\ref{constraint1}) using the trace $T=-\rho+3p(\rho)$ and the
conservation equation $\dot{\rho}+3H(\rho+p)=0$, could
in principle leads to a differential equation for
the function $f(R,T)$. For the expression
(\ref{fRT function}) it reads
\begin{align}\label{conservation}
\Big{(} \dot{h_{1}'}\rho^{\textrm{(b)}}+
\frac{1}{2}h_{1}'\dot{\rho}^{\textrm{(b)}} \Big{)}+
\Big{[} \dot{h_{2}'}\big{(}\rho^{\textrm{(G)}}+
p^{\textrm{(G)}}\big{)}-\frac{1}{2}h_{2}'\big{(}
\dot{p}^{\textrm{(G)}}-\dot{\rho}^{\textrm{(G)}}\big{)}\Big{]}=0.
\end{align}
Therefore, assuming that the two matters do not interact with each
other, we can obtain
\begin{align}
&\dot{h_{1}'}\rho^{\textrm{(b)}}+
\frac{1}{2}h_{1}'\dot{\rho}^{\textrm{(b)}}=0, \label{DE-m}\\
&\dot{h_{2}'}\big{(}\rho^{\textrm{(G)}}+p^{\textrm{(G)}}\big{)}-
\frac{1}{2}h_{2}'\big{(}\dot{p}^{\textrm{(G)}}-
\dot{\rho}^{\textrm{(G)}}\big{)}=0.\label{DE-G}
\end{align}
Since the signature of metric (\ref{metricFRW}) implies
 $\rho^{\textrm{(b)}}=-T^{\textrm{(b)}}$ for the
pressureless matter, equation (\ref{DE-m}) gives
\begin{align}\label{conservedfunction-m}
h_{1}\big{(}T^{\textrm{(b)}}\big{)}=C_{1}^{\textrm{(b)}}
\sqrt{-T^{\textrm{(b)}}}+C_{2}^{\textrm{(b)}}.
\end{align}
where $C_{1}^{\textrm{(b)}}$ and $C_{2}^{\textrm{(b)}}$ are
constants of integration. Now we try to solve equation (\ref{DE-G})
for GCG. The EoS for GCG is written as
\begin{align}\label{chaplygin-p}
P^{\textrm{(G)}}=-\frac{A}{\rho^{\textrm{(G)}\alpha}}.
\end{align}
Equation (\ref{chaplygin-p}) together with the equation
of the EMT conservation $\dot{\rho}^{\textrm{(G)}}+
3H(\rho^{\textrm{(G)}}+p^{\textrm{(G)}})=0$ lead to the
following solution for the matter density of GCG
\begin{align}\label{chaplygin-rho1}
\rho^{\textrm{(G)}}=(A+Ba^{-3(1+\alpha)})^{\frac{1}{1+\alpha}},
\end{align}
where $A$, $\alpha$ and $B$ are some constant which have to satisfy the conditions
$A>0$ and $\alpha\geq0$~\footnote{These restrictions come from the GCG sound speed
considerations. See, e.g., Ref.~\cite{CG4}.}. Setting $\rho^{\textrm{(G)}}_
{0}\equiv\rho^{\textrm{(G)}}(a=1)$, we can rewrite (\ref{chaplygin-rho1})
in a suitable form as
\begin{align}\label{chaplygin-rho2}
\rho^{\textrm{(G)}}=\rho^{\textrm{(G)}}_{0}\left(K+(1-K)a^{-3(1+\alpha)}\right)
^{\frac{1}{1+\alpha}},
\end{align}
where $a$ is the scale factor and $K\equiv A/\rho^{\textrm{(G)}(1+\alpha)}_{0}$.
For later applications we rewrite (\ref{chaplygin-rho2}) as
\begin{align}\label{chaplygin-rho20}
\rho^{\textrm{(G)}}=\rho^{\textrm{(G)}}_{0}u(a;\alpha,K),
\end{align}
where
\begin{align}\label{chaplygin-rho201}
u(a;\alpha,K)\equiv \left(K+(1-K)a^{-3(1+\alpha)}\right)
^{\frac{1}{1+\alpha}},
\end{align}
where $u(1;\alpha,K)=1$. Note that, the argument $(a;\alpha,K)$ denotes
the variation with respect to the scale factor for constant values of $\alpha$
and $K$. Substituting equation (\ref{chaplygin-p}) in (\ref{DE-G}) together with
using $\dot{\rho}^{\textrm{(G)}}+3H(\rho^{\textrm{(G)}}+ p^{\textrm{(G)}})=0$
and $T^{\textrm{(G)}}=-\rho^{\textrm{(G)}}+3p^{\textrm{(G)}}$, we get
\begin{align}\label{conservedfunction-G1}
2\frac{h_{2}''}{h_{2}'}=\frac{\rho^{\textrm{(G)}}+
\alpha p^{\textrm{(G)}}}{(\rho^{\textrm{(G)}}+
3\alpha p^{\textrm{(G)}})(\rho^{\textrm{(G)}}+p^{\textrm{(G)}})}.
\end{align}
The left hand side of equation (\ref{conservedfunction-G1})
is in terms of $T^{\textrm{(G)}}$ while the right hand
side is a function of $\rho^{\textrm{(G)}}$ and  $p^{\textrm{(G)}}$,
which shows that it is not a closed deferential equation.
To complete it, the trace of EMT of
GCG can be rewritten as the following form
\begin{align}\label{trace-G}
-\rho^{\textrm{(G)}}-3\frac{A}{\rho^{\textrm{(G)}~\alpha}}=T^{\textrm{(G)}}.
\end{align}
By solving equation (\ref{trace-G}), in principle, we can obtain
the matter density $\rho^{\textrm{(G)}}$ in terms of the GCG trace,
and then, substitute it into equations (\ref{chaplygin-p}) and
(\ref{conservedfunction-G1}) to get a closed differential equation
in terms of the pure trace $T^{\textrm{(G)}}$. Unfortunately,
this is not a straightforward calculation, since equation
(\ref{trace-G}) admits a lot of roots that depend on the
value $\alpha$ (equation (\ref{trace-G}) can be problematic when
non-integer values of $\alpha$ are included as well). Furthermore,
the solutions can get complicated forms such that the differential
equation (\ref{conservedfunction-G1}) may not be solved. However,
for the case of the standard Chaplygin gas (SCG), namely,
$\alpha=1$ the solutions are tractable and for the other values
we use the approximation methods. Equation (\ref{trace-G}) for
$\alpha=1$ has the following solutions
\begin{align}\label{rho-G-exact}
\rho^{\textrm{(S)}}_{\pm}=\frac{1}{2} \left(-T^{\textrm{(S)}}
\pm\sqrt{-12 A+T^{\textrm{(S)}~2}}\right),
\end{align}
where superscript ``S" stands for SCG.
Since, $\rho^{\textrm{(S)}}_{-}$ can get negative values and thus
violates the weak energy condition (WEC),
we discard it as a non-physical solution. We then
consider the solution $\rho^{\textrm{(S)}}_{+}$.
By setting $\alpha=1$ in equation
(\ref{conservedfunction-G1}), and using (\ref{rho-G-exact})
we have
\begin{align}\label{diff-1}
2\frac{h_{2,+}''}{h_{2,+}'}=\frac{1}{(2\rho_{+}^{\textrm{(S)}}+T^{\textrm{(S)}})}
=\frac{1}{\sqrt{-12 A+T^{\textrm{(S)}~2}}},
\end{align}
and, we finally obtain the following differential equation
\begin{align}\label{diff-2}
2\sqrt{-12 A+T^{\textrm{(S)}~2}}h_{2,+}''-h_{2,+}'=0,
\end{align}
where the solution $h_{2,+}$ corresponds to $\rho_{+}^{\textrm{(S)}}$.
The solutions of the above differential equation are obtained as
\begin{align}\label{conservedfunction-M1}
h_{2,+}\Big{(}T^{\textrm{(S)}}\Big{)}= -\frac{2}{3} C_{1+}^{\textrm{(S)}}
\left(-2 T^{\textrm{(S)}}+\sqrt{-12 A+T^{\textrm{(S)}~2}}\right)
\sqrt{T^{\textrm{(S)}}+\sqrt{-12 A+T^{\textrm{(S)}~2}}}+C_{2+}^{\textrm{(S)}},
\end{align}
where $C_{1+}^{\textrm{(S)}}$ and $C_{2+}^{\textrm{(S)}}$ are
constants of integration. As a result, in $f(R,T)$ theory of gravity,
for the non-interacting baryonic matter and SCG, the conservation
of EMT enforces us to use the following function
\begin{align}\label{fRT-I}
&f^{\textrm{(I)}}\Big{(}R, T^{\textrm{(b, S)}}\Big{)}=R+C_{1}^{\textrm{(b)}}
\sqrt{-T^{\textrm{(b)}}}\nonumber\\
&-\frac{2}{3} C_{1}^{\textrm{(S)}}
\left(-2 T^{\textrm{(S)}}+\sqrt{-12 A+T^{\textrm{(S)}~2}}\right)
\sqrt{T^{\textrm{(S)}}+\sqrt{-12 A+T^{\textrm{(S)}~2}}}+\Lambda^{\textrm{(b, S)}},
\end{align}
where we have restored the argument of function $f$ for clarification
and dropped the sign ``$+$". Also, we have added the superscript I
for later applications and $\Lambda^{\textrm{(b,S)}}\equiv
C_{2}^{\textrm{(b)}}+C_{2}^{\textrm{(S)}}$. Nevertheless, we
consider two another model; the models which use GCG in two
extreme situations, i.e., when $p^{\textrm{(G)}}\gg\rho^
{\textrm{(G)}}$ and $\rho^{\textrm{(G)}}\gg p^{\textrm{(G)}}$.
These cases will be considered for arbitrary values of $\alpha$.
In either cases we approximate $T^{\textrm{(G)}}_{p}\simeq
3p^{\textrm{(G)}}=-3|p^{\textrm{(G)}}|$ and $T^{\textrm{(G)}}_
{\textrm{$\rho$}}\simeq -\rho^{\textrm{(G)}}$. In these cases,
using equation (\ref{DE-G}), the related differential equations
are given as
\begin{align}\label{diff-app}
&2T^{\textrm{(G)}}_{p} h_{2,p}''\Big{(}T^{\textrm{(G)}}_{p}\Big{)}-
h_{2,p}'\Big{(}T^{\textrm{(G)}}_{p}\Big{)}\simeq0,\\
&2T^{\textrm{(G)}}_{\rho} h_{2,\rho}''\Big{(}T^{\textrm{(G)}}_{\rho}\Big{)}+
h_{2,\rho}'\Big{(}T^{\textrm{(G)}}_{\rho}\Big{)}\simeq0,
\end{align}
for which the solutions are found as
\begin{align}\label{final function-app1}
h_{2,p}\Big{(}T^{\textrm{(G)}}_{p}\Big{)}=\frac{2}{3} C^{\textrm{(G)}}_{1p}
\left(-T^{\textrm{(G)}}\right)~^{3/2}+C^{\textrm{(G)}}_{2p},
\end{align}
and
\begin{align}\label{final function-app2}
h_{2,\rho}\Big{(}T^{\textrm{(G)}}_{\rho}\Big{)}=2 C^{\textrm{(G)}}_{1\rho}
\sqrt{-T^{\textrm{(G)}}}+C^{\textrm{(G)}}_{2\rho},
\end{align}
where we have restored again the arguments and labeled the
equations and solutions by $p$ and $\rho$ that denote the
either extreme cases and $C^{\textrm{(G)}}_{ip}$ and
$C^{\textrm{(G)}}_{i\rho}$ with $i=1,2$ are integral
constants. Thus, we have two another forms for $f(R,T)$
function, i.e.,
\begin{align}\label{fRT-II}
f^{\textrm{(II)}}\Big{(}R, T^{\textrm{(b, G)}}\Big{)}=R+C_{1}^{\textrm{(b)}}
\sqrt{-T^{\textrm{(b)}}}+\frac{2}{3} C^{\textrm{(G)}}_{1p}
\left(-T^{\textrm{(G)}}\right)~^{3/2}+\Lambda^{\textrm{(b, G)}}_{p},
\end{align}
for $\mid p^{\textrm{(G)}}\mid \gg\rho^{\textrm{(G)}}$, where
$\Lambda^{\textrm{(b, G)}}_{p}\equiv C_{2}^{\textrm{(b)}}
+C_{2p}^{\textrm{(G)}}$ and
\begin{align}\label{fRT-III}
f^{\textrm{(III)}}\Big{(}R, T^{\textrm{(b, G)}}\Big{)}=R+C_{1}^{\textrm{(b)}}
\sqrt{-T^{\textrm{(b)}}}+2 C^{\textrm{(G)}}_{1\rho}
\sqrt{-T^{\textrm{(G)}}}+\Lambda^{\textrm{(b, G)}}_{\rho},
\end{align}
for $\rho^{\textrm{(G)}}\gg \mid p^{\textrm{(G)}}\mid$,
where $\Lambda^{\textrm{(b, G)}}_{\rho}\equiv C_{2}
^{\textrm{(b)}}+C_{2\rho}^{\textrm{(G)}}$. Note that,
there are two sets of constants labeled by the numbers $1$
and $2$ which determine the degree of involved differential equations.

Therefore, we have three classes of non-interacting models including
pressureless baryonic and three forms of CG; in the first one we
have SCG represented by the function (\ref{fRT-I}), the second
one uses GCG when the matter density is negligible in comparison
with the pressure as shown by the function (\ref{fRT-II}), and in the
last case, the GCG matter density is dominant which is determined
by function (\ref{fRT-III}).

Note that, from equations (\ref{chaplygin-p}) and (\ref{chaplygin-rho2}),
the EoS parameter $w^{(\textrm{GCG})}$ for GCG can be obtained. At the
early time ($a\rightarrow0$), we have $w^{(\textrm{GCG})}\rightarrow0$,
and at the late time ($a\rightarrow\infty$), $w^{(\textrm{GCG})}\rightarrow-1$.
This means that the GCG in the early time behaves like CDM and in the late time
like DE. Therefore, we assume that DE and CDM are unified by GCG model.
Besides, we have the baryonic matter indicated by superscript ``b" which
does not interact with GCG component.

In the next section, we present the statefinder parameters and obtain them
for these three models, and also get the deceleration parameter and define
the effective EoS, as well. Henceforth, we will call these as, models I, II and III.
\section{Statefinder parameters and the related definitions}\label{eq-def}
In this section, we present the definitions of the statefinder
parameters and calculate them for models I, II and III.
These parameters are important to discuss the cosmological
aspects of models which are introduced in Refs.~\cite{stfi1,stfi2},
originally. The statefinder parameters are defined
via the following pairs of parameters~\cite{stfi1,stfi2}
\begin{align}
&r\equiv\frac{\dddot{a}}{a}H^{-3},\label{rr}\\
&s\equiv\frac{r-1}{3(q-1/2)}.\label{ss}
\end{align}
Models that have the same present values of the Hubble
parameter $H_{0}$ and the deceleration parameter $q_{0}$,
can be discriminated from each other by these parameters. These parameters include
the third time derivative of the scale factor (in spite of the Hubble
parameter which includes the first time derivative
and the deceleration parameter which includes the second time derivative
of the scale factor) and can be used to distinguish the different DE
models. The deceleration parameter is defined as the first time
derivative of the Hubble parameter, i.e.,
\begin{align}\label{deceleration}
q\equiv-\frac{\dot{H}}{H^2}-1,
\end{align}
which in terms of the normalized Hubble parameter
$E(a)=H/H_{0}$, it can be rewritten as
\begin{align}\label{deceleration-E}
q=-\frac{1}{E}\frac{dE}{dN}-1,
\end{align}
where $N\equiv \ln{a}$. Using the definition (\ref{deceleration}),
the statefinder parameter $r$ can be simply calculated as
\begin{align}\label{r-q}
r=q(1+2q)-\frac{dq}{dN},
\end{align}
whereby substituting (\ref{deceleration-E}) in (\ref{r-q}) gives
\begin{align}\label{r-E}
r=\frac{1}{E}\frac{d^{2}E}{dN^{2}}+\frac{1}{E^{2}}\left(\frac{dE}{dN}\right)^{2}+\frac{3}{E}\frac{dE}{dN}+1.
\end{align}
One can use equations (\ref{deceleration-E}) and (\ref{r-E}) to obtain the
statefinder parameter $s$ from definition (\ref{ss}).
According to the definition of the normalized Hubble parameter,
and also using (\ref{Hubble}) for models (\ref{fRT-I}), (\ref{fRT-II})
and (\ref{fRT-III}) we obtain
\begin{align}\label{EI}
E^{(\textrm{I})}(a;1,K)=\Bigg{[}\Omega_{0}^{\textrm{(b)}}\left(1-\mathbf{m}a^{3/2}\right)a^{-3}+
\Omega_{0}^{\textrm{(S)}}u(a;1,K)\left(1-\sqrt{24} \mathbf{n^{(\textrm{I})}}
\left(\frac{u(a;1,K)}{K}\right)^{-3/2}\right)\Bigg{]}^{\frac{1}{2}},
\end{align}
\begin{align}\label{EII}
E^{(\textrm{II})}(a;\alpha,K)=\Bigg{[}\Omega_{0}^{\textrm{(b)}}\left(1-\mathbf{m}a^{3/2}\right)a^{-3}+
\Omega_{0}^{\textrm{(G)}}u(a;\alpha,K)\left(1-\sqrt{12} \mathbf{n^{(\textrm{II})}}K^{\frac{3}{2}}
u(a;\alpha,K)^{-\frac{3\alpha+2}{2}}\right)\Bigg{]}^{\frac{1}{2}},
\end{align}
and
\begin{align}\label{EIII}
E^{(\textrm{III})}(a;\alpha,K)=\Bigg{[}\Omega_{0}^{\textrm{(b)}}\left(1-\mathbf{m}a^{3/2}\right)a^{-3}+
\Omega_{0}^{\textrm{(G)}}u(a;\alpha,K)\left(1-2 \mathbf{n^{(\textrm{III})}}u(a;\alpha,K)
^{-\frac{1}{2}}\right)\Bigg{]}^{\frac{1}{2}},
\end{align}
where we have defined the density parameters for the two type
of fluids and some dimensionless parameters, as
\begin{align}\label{dimensionless}
&\Omega_{0}^{\textrm{(b)}}=\frac{8\pi G \rho_{0}^{\textrm{(b)}}}{3H_{0}^{2}},~~~~~~~~
\Omega_{0}^{\textrm{(S/G)}}=\frac{8\pi G \rho_{0}^{\textrm{(S/G)}}}{3H_{0}^{2}},~~~~~~~~~
\mathbf{m}=\frac{\rho^{\textrm{(b)}~-1/2}_{0}C_{1}^{\textrm{(b)}}}{8\pi G},\nonumber\\
&\mathbf{n^{(\textrm{I})}}=\frac{\rho^{\textrm{(S)}~1/2}_{0}C_{1}^{\textrm{(S)}}}{8\pi G},~~~~~
\mathbf{n^{(\textrm{II})}}=\frac{\rho^{\textrm{(G)}~1/2}_{0}C_{1p}^{\textrm{(G)}}}{8\pi G},~~~~~
\mathbf{n^{(\textrm{III})}}=\frac{\rho^{\textrm{(G)}~-1/2}_{0}C_{1\rho}^{\textrm{(G)}}}{8\pi G}.\
\end{align}
Hence, each model includes some parameters; the space parameters of model
I are ($\mathbf{m},\mathbf{n^{(\textrm{I})}},K$), those of
model II are ($\mathbf{m},\mathbf{n^{(\textrm{II})}},K,\alpha$)
and for model III we have ($\mathbf{m},\mathbf{n^{(\textrm{III})}},K,\alpha$).
These parameters play the role of some tuners for these models which can be set for a
physically interesting situation. Substituting equations (\ref{EI}),
(\ref{EII}) and (\ref{EIII}) in equation (\ref{deceleration-E}) gives the
related deceleration parameters as
\begin{align}\label{qI}
q^{(\textrm{I})}=\frac{3}{4E^{(\textrm{I}),2}}\left[\Omega _{0}^{\textrm{(b)}}
(2-\mathbf{m}a^{3/2})a^{-3}+2 \Omega_{0}^{\textrm{(S)}} (1-K)uv
\left(1+\sqrt{6} \mathbf{n^{(\textrm{I})}} (u/K)^{-3/2}\right)\right]-1,
\end{align}
\begin{align}\label{qII}
q^{(\textrm{II})}=\frac{3}{4E^{(\textrm{II}),2}}
\left[\Omega_{0}^{\textrm{(b)}}(2-\mathbf{m}a^{3/2})
a^{-3}+2\Omega_{0}^{\textrm{(G)}}(1-K)v\left(u+\sqrt{27}
\alpha\mathbf{n^{(\textrm{II})}} \left(\frac{u^{\alpha}}{K}
\right)^{-3/2}\right)\right]-1,
\end{align}
and
\begin{align}\label{qIII}
q^{(\textrm{III})}=\frac{3}{4E^{(\textrm{III}),2}}
\left[\Omega _{0}^{\textrm{(b)}}(2-\mathbf{m}a^{3/2})a^{-3}+
2\Omega_{0}^{\textrm{(G)}} (1-K)uv\left(1-
\mathbf{n^{(\textrm{III})}} u^{-\frac{1}{2}}\right) \right]-1,
\end{align}
for models I, II and III, respectively. It is convenient to define an effective
EoS in the higher order gravity (specially in $f(R)$ gravity) to explain
the present accelerated expansion of the Universe through the effects of the geometrically modified
gravity terms in the Friedmann equations\footnote{For example, see Ref.~\cite{fR}
and references therein.}. Here, we have not taken into account such modifications in the geometrical sector of
the action. However, there is a non-standard interaction between cosmological fluids and normal
geometrical sector, which is the general behavior of $f(R,T)$ gravity.
The effective EoS parameter $w^{(\textrm{eff})}$ is defined as
$w^{(\textrm{eff})}=-1-2\dot{H}/3H^{2}$~\footnote{See Refs.~\cite{fRT7,fRT8}}, whereby using the definition
(\ref{deceleration}) we get,
\begin{align}\label{EoS}
w^{(\textrm{eff})}=\frac{1}{3}(2q-1).
\end{align}
The effective EoS can be obtained from equations (\ref{qI}),
(\ref{qII}), (\ref{qIII}) and (\ref{EoS}) for the three models.
Using equations (\ref{r-E}), (\ref{EI}), (\ref{EII}) and (\ref{EIII}) the statefinder parameters $r$
can be calculated as
\begin{align}\label{rI}
&r^{(\textrm{I})}=\frac{9}{8E^{(\textrm{I}),2}}
\Bigg\{\Omega_{0}^{\textrm{(b)}}\mathbf{m}a^{-3/2}+
\Omega_{0}^{\textrm{(S)}}(1-K)u^{-1}v
\bigg{[}4 K+\Bigg.\nonumber\\
&\Bigg.\sqrt{24}\mathbf{n^{(\textrm{I})}}\Big{(}2-5(1-K)v
\Big{)}(u/K)^{-3/2}u^{2}\bigg{]}\Bigg\}+1,
\end{align}
\begin{align}\label{rII}
&r^{(\textrm{II})}=\frac{9}{8E^{(\textrm{II}),2}}\Bigg
\{\Omega _{0}^{\textrm{(b)}}\mathbf{m} a^{-3/2}+2 \alpha
\Omega_{0}^{\textrm{(G)}}(1-K)v\bigg{[}2
\Big{(}1-(1-K)v\Big{)}u+\bigg.\Bigg.\nonumber\\
&\Bigg.\bigg.\sqrt{27}\mathbf{n^{(\textrm{II})}}
\Big{(}2\alpha -(1-K)(2+5\alpha)v \Big{)}(\frac{u^{\alpha}}{K})^{-3/2}\bigg{]}\Bigg\}+1,
\end{align}
\begin{align}\label{rIII}
&r^{(\textrm{III})}=\frac{9}{8E^{(\textrm{III}),2}}
\Bigg\{\Omega _{0}^{\textrm{(b)}}\mathbf{m}a^{-3/2}+2
\Omega_{0}^{\textrm{(G)}}(1-K)v u\bigg{[}2\alpha
\Big{(}1-(1-K)v\Big{)}-\bigg.\nonumber\\
&\bigg.\mathbf{n^{(\textrm{III})}}\Big{(}2
\alpha-(1-K)(1+2\alpha)v\Big{)}u^{-\frac{1}{2}}\bigg{]}\Bigg\}+1,
\end{align}
and the statefinder $s$ for three models can be computed from definition
(\ref{ss}), using the above equations for the deceleration parameters
$q^{(\textrm{I})}, q^{(\textrm{II})}, q^{(\textrm{III})}$
and the statefinder parameters $r^{(\textrm{I})}, r^{(\textrm{II})}, r^{(\textrm{III})}$, as
\begin{align}\label{sI}
s^{(\textrm{I})}=\frac{1}{2}\frac{\Omega _{0}^{\textrm{(b)}}
\mathbf{m}a^{-3/2}+\Omega_{0}^{\textrm{(S)}}(1-K)u^{-1}v
\bigg{[}4K+\sqrt{24}\mathbf{n^{(\textrm{I})}}\Big{(}2-5(1-K)v
\Big{)}(u/K)^{-3/2}u^{2}\bigg{]}}{\Omega_{0}^{\textrm{(b)}}
\mathbf{m}a^{-3/2}-\Omega_{0}^{\textrm{(S)}}u^{-1}
\bigg{[}2K-\sqrt{24}\mathbf{n^{(\textrm{I})}}
\Big{(}2+(1-K)v\Big{)}(u/K)^{-3/2}u^{2}\bigg{]}},
\end{align}
\begin{align}\label{sII}
s^{(\textrm{II})}=\frac{1}{2}\frac{\Omega _{0}^{\textrm{(b)}}
\mathbf{m} a^{-3/2}+2 \alpha\Omega_{0}^{\textrm{(G)}}(1-K)v
\bigg{[}2 \Big{(}1-(1-K)v\Big{)}u+\sqrt{27}\mathbf{n^{(\textrm{II})}}
\Big{(}2\alpha-(1-K)(2+5\alpha)v
\Big{)}(\frac{u^{\alpha}}{K})^{-\frac{3}{2}}
\bigg{]}}{\Omega_{0}^{\textrm{(b)}}\mathbf{m} a^{-3/2}-
\Omega_{0}^{\textrm{(G)}}\bigg{[}2 \Big{(}1-\left(1-K\right)v
\Big{)}u-\sqrt{12} \mathbf{n^{(\textrm{II})}}\Big{(}2+3 \alpha(1-K)v
\Big{)}(\frac{u^{\alpha }}{K})^{-\frac{3}{2}}\bigg{]}},
\end{align}
\begin{align}\label{sIII}
s^{(\textrm{III})}=\frac{1}{2}\frac{\Omega _{0}^{\textrm{(b)}}
\mathbf{m}a^{-3/2}+2\Omega_{0}^{\textrm{(G)}}(1-K)v u\bigg{[}2\alpha
\Big{(}1-(1-K)v\Big{)}-\mathbf{n^{(\textrm{III})}}
\Big{(}2\alpha-(1-K)(1+2\alpha)v\Big{)}u^{-\frac{1}{2}}
\bigg{]}}{\Omega _{0}^{\textrm{(b)}}\mathbf{m} a^{-3/2}-2u
\Omega_{0}^{\textrm{(G)}}\bigg{[}1-(1-K)v-\mathbf{n^{(\textrm{III})}}
\Big{(}2-(1-K)v\Big{)}u^{-\frac{1}{2}}\bigg{]}},
\end{align}
where we have defined $v(a;\alpha,K)=[a^{3}u(a;\alpha,K)]^{-(1+\alpha)}$, and also
dropped the arguments $(a;1,K)$ and $(a;\alpha,K)$ from the quantities of model
I and the two other ones, respectively. Note that, the dimensionless parameters introduced in
(\ref{dimensionless}) are not independent. Applying the present value $E_{0}^{(\textrm{I})}(1;\alpha,K)=1$,
and also the same for the other models, we get
\begin{align}
&\Omega_{0}^{\textrm{(S)}}=\frac{1-(1-\mathbf{m})\Omega_{0}^{\textrm{(b)}}}
{1-\sqrt{12}\mathbf{n}^{\textrm{(I)}}K^{\frac{3}{2}}}~~~~~~~~~~~~\mbox{for model I},\label{rel1}\\
&\Omega_{0}^{\textrm{(G)}}=\frac{1-(1-\mathbf{m})\Omega_{0}^{\textrm{(b)}}}
{1-\sqrt{24}\mathbf{n}^{\textrm{(II)}}K^{\frac{3}{2}}}~~~~~~~~~~~~\mbox{for model II},\label{rel2}\\
&\Omega_{0}^{\textrm{(G)}}=\frac{1-(1-\mathbf{m})\Omega_{0}^{\textrm{(b)}}}
{1-2\mathbf{n}^{\textrm{(III)}}}~~~~~~~~~~~~\mbox{for model III}.\label{rel3}
\end{align}
Setting $\mathbf{m}=0$ and $\mathbf{n}$'s$=0$, in equations
(\ref{qI})-(\ref{sIII}) and also relations (\ref{rel1})-(\ref{rel3})
gives the corresponding equations for model $f(R,T)=R$, i.e. the GR case.
These coupling constants are responsible for all deviations of the above equations
from the equivalent forms in GR. In the next section we study these deviations
for three models I, II and III, numerically. That is, we investigate the
cosmological consequences of equations (\ref{qI})-(\ref{sIII}) through the statefinder diagnosis.
\section{Statefinder diagnosis and numerical considerations}\label{numerical}
In this section we consider equations (\ref{qI})-(\ref{sIII}) and
extract their cosmological consequences via the statefinder diagnosis.
These three models depend on the coupling constants $\mathbf{m}$ and
$\mathbf{n}$'s, and also parameter $K$ for model I, and
$K,\alpha$ for the two other ones. Generally, these constants
can lead to different models for different values, and therefore, different
cosmological history for the universe. In the following subsections we
consider each model, in turn.

Note that, in each forthcoming diagram, the trajectories of cosmological parameters
for $\mathbf{m}$, $\mathbf{n}$'s=0 correspond to the GR background.
Namely, one can compare the $f(R,T)$ gravity results with the corresponding GR ones
in each model I, II and III, using these plots.

Later, in Sec.~\ref{test}, we compare
the predicted present values $s_{0}$ and $j_{0}$ by our models (which have been obtained for the
best fit values of models parameters), to some observational
measurements. The gold sample supernova type Ia which gives $1.65<r_{0}<3.97$~\cite{riess},
the SNLS supernova type Ia data set which results in $0.11<r_{0}<2.69$~\cite{astier} and
X-ray galaxy clusters analysis which gives $-1.49<r_{0}<3.06$~\cite{rapetti} have been used. Future
measurements like SNIa data from the Large Synaptic Survey Telescope (LSST),
the Supernovae Acceleration Probe (SNAP) and X-ray cluster data from
Constellation-X may provide better constraints on the statefinder parameters.
The baryon oscillation experiment which is applied in galaxy redshift surveys
for high-redshift ranges can also make tighter constraint on these parameters.
\subsection{model $f^{\textrm{(I)}}\big{(}R, T^{\textrm{(b, S)}}\big{)}$}\label{Model I}
\begin{figure}[h]
\centering
\epsfig{figure=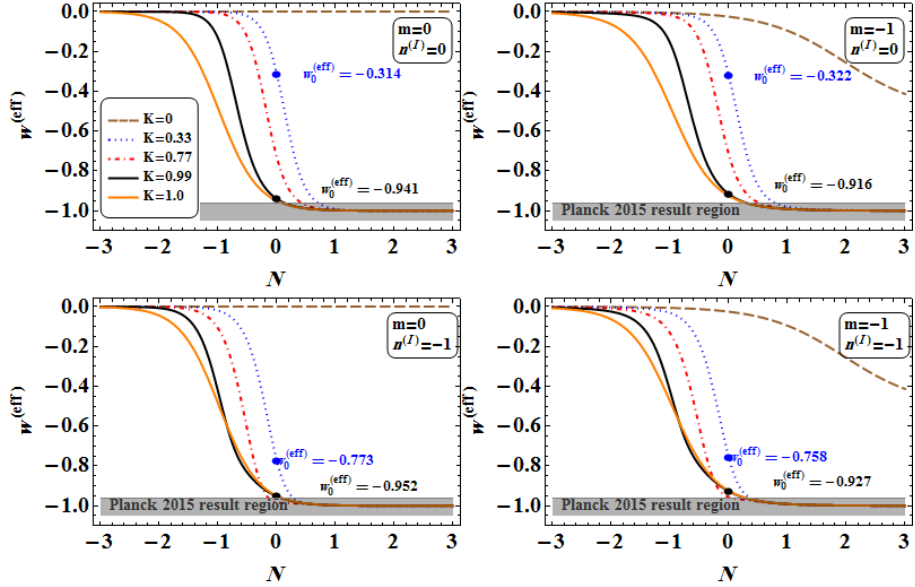,width=12cm}
\caption{(color online). \footnotesize {The cosmological evolution
of the effective EoS parameter of model I versus
$N\equiv\ln a$. The curves are plotted for $K=0$ (which in this case we have
CDM in addition to the baryonic one) in brown dashed line, blue dotted line for
$K=0.33$, red dot-dashed line for case $K=0.77$, black
solid line for $K=0.99$ and orange solid line for $K=1.00$. Small solid
circles determine the present values. The valid range of values which has been reported by Planck 2015
measurements, ($-1.051\leq w_{0, \textrm{Planck}}^{(\textrm{eff})}\leq-0.961$) is also
indicated by the gray region. All diagrams are
drawn for the present baryonic matter density
$\Omega_{0}^{\textrm{(b)}}=0.05$. Since, positive
values of parameters $\mathbf{m}$ and $\mathbf{n}^{(\textrm {I})}$ lead to
some abnormality, only negative values are discussed. The
curve of $K=1.00$ is not flat because of the effect of the baryonic
matter density. Zero value for the coupling constants $\mathbf{m}$
and $\mathbf{n}^{(\textrm {I})}$ reflects the GR background. Columnar view shows that
decreasing the value of $\mathbf{n}^{(\textrm {I})}$ leads to decreasing the present
values. However, lateral view demonstrates that decreasing $\mathbf{m}$
leads to an increase for $K>0.5$ and a decrease for $K<0.5$. An
orthogonal look shows that the overall effect of switching on both
parameter leads to more observationally accepted values for $K<0.5$. For larger
values of $\mathbf{m}$ and $\mathbf{n}^{(\textrm {I})}$} we get the better results.}
\label{wi}
\end{figure}

Using equations (\ref{qI}) and (\ref{EoS}), the cosmological evolution
of the EoS parameter for this model is illustrated in Fig.~\ref{wi},
for specific values $K=0,~0.33,~0.77$ and $0.99$ with
$\mathbf{m},\mathbf{n^{(\textrm{I})}}=-1$. Moreover, we have
plotted the corresponding diagrams for the standard case
$\mathbf{m}=0, \mathbf{n^{(\textrm{I})}}=0$ (i.e., for GR,
as we have before mentioned) in order to indicate possible
deviations. The zero value for $K$, corresponds
to the model with a pressureless matter (see equation
(\ref{chaplygin-rho2})), and the other values determine
different cosmological scenarios. The predicted present
value $w_{0}^{\textrm{(eff)}}$, is indicated by a solid
circle on each curve. Also, we have illustrated
observational range for the values of EoS parameter of DE
which has been reported by Planck 2015 measurements~\cite{Planck2},
($-1.051\leq w_{0, \textrm{Planck}}^{(\textrm{eff})}\leq-0.961$),
in Fig.~\ref{wi} by a gray region.
Brown dashed line has been used for case $K=0$, blue dotted
one for $K=0.33$, red dot-dashed curve shows the
case $K=0.77$, a black line presents the models with $K=0.99$ and
an orange one for $K=1$\footnote{In the subsequent discussions and subsections we always use these options
to show the related plots for $K=0,0.33,0.77,0.99$.}.
Since, there are generally some divergences in the cases with
positive values of $\mathbf{n^{(\textrm{I})}}$ and $\mathbf{m}$ (which occur in a
certain value of $N$, for some values of $K$), the evolution of
the related EoS parameter is not depicted in Fig.~\ref{wi}.
Physically, this divergence causes a total increase in the value of the
deceleration parameter after the matter dominated era\footnote{Note that, for this reason, we do not consider the
other cosmological parameters for this model and also for two other ones
for positive valued parameters $\mathbf{m},\mathbf{n}$'s.}. Hence,
positive values for the coupling constants $\mathbf{m},
\mathbf{n^{(\textrm{I})}}$, can lead to non-physical behaviors and we
discard such cases as the less physically interesting
models. However, one may find some curves without divergence for positive values.
For example, for $\mathbf{m}=2,~\mathbf{n^{(\textrm{I})}}=-2.3$
and $K=0.32$ we can obtain the acceptable value $w^{(\textrm{eff})}\simeq-1.005$
consistent with the recent results~\cite{WMAP5,Planck2}. Plots in Fig.~\ref{wi} show that the value of the effective EoS
is much more near to its observationally obtained values,
$w^{\textrm{(obs)}}\simeq-1$, for $\mathbf{n^{(\textrm{I})}}\neq0$.
Decreasing the parameter $\mathbf{n^{(\textrm{I})}}$ for $\mathbf{m}=0$, leads to smaller values
of $w^{\textrm{(eff)}}$ for all values of $K$. By contrast,
decreasing in the parameter $\mathbf{m}$ for $\mathbf{n^{(\textrm{I})}}=0$,
gives smaller values for $K<1/2$ and larger values for $K>1/2$. To understand the reason of
this dual behavior, we calculate the present value of $w^{(\textrm{eff})}$ for
$\mathbf{n^{(\textrm{I})}}=0$, as
\begin{align}\label{wI0}
w_{0}^{(\textrm{eff})}=K(\Omega_{0}^{(\textrm{b})}-1) -
\frac{1}{2}(2K-1)\mathbf{m}\Omega_{0}^{(\textrm{b})},~~~~~~\mathbf{n^{(\textrm{I})}}=0,~~~~~~N=0.
\end{align}
As we see, for special case $K=1/2$, the effective EoS is
independent of the parameter $\mathbf{m}$ and for a given $\mathbf{m}$, the
value of $w_{0}^{(\textrm{eff})}$ increases for $K>1/2$ and
decreases for $K<1/2$. Calculations show that when $\mathbf{n^{(\textrm{I})}}\neq0$,
the mutual behavior is absent and in these situations a decrease in $\mathbf{m}$ leads
to an increase in $w_{0}^{(\textrm{eff})}$ for all values of $K$ (compare lower diagrams
in Fig.~\ref{wi}). Therefore, in
lower right diagram we see a total decrease in the value of $w_{0}^{(\textrm{eff})}$
(a decrease due to the effect of $\mathbf{n^{(\textrm{I})}}\neq0$ and an increase due to
the effect of $\mathbf{m}$).
The value of $w_{0}^{(\textrm{eff})}$ vanishes
in the early times ($N<0$ which corresponds to $a<1$), and goes to $-1$ in the
late times except for the case $K=0$.

\begin{figure}[h]
\epsfig{figure=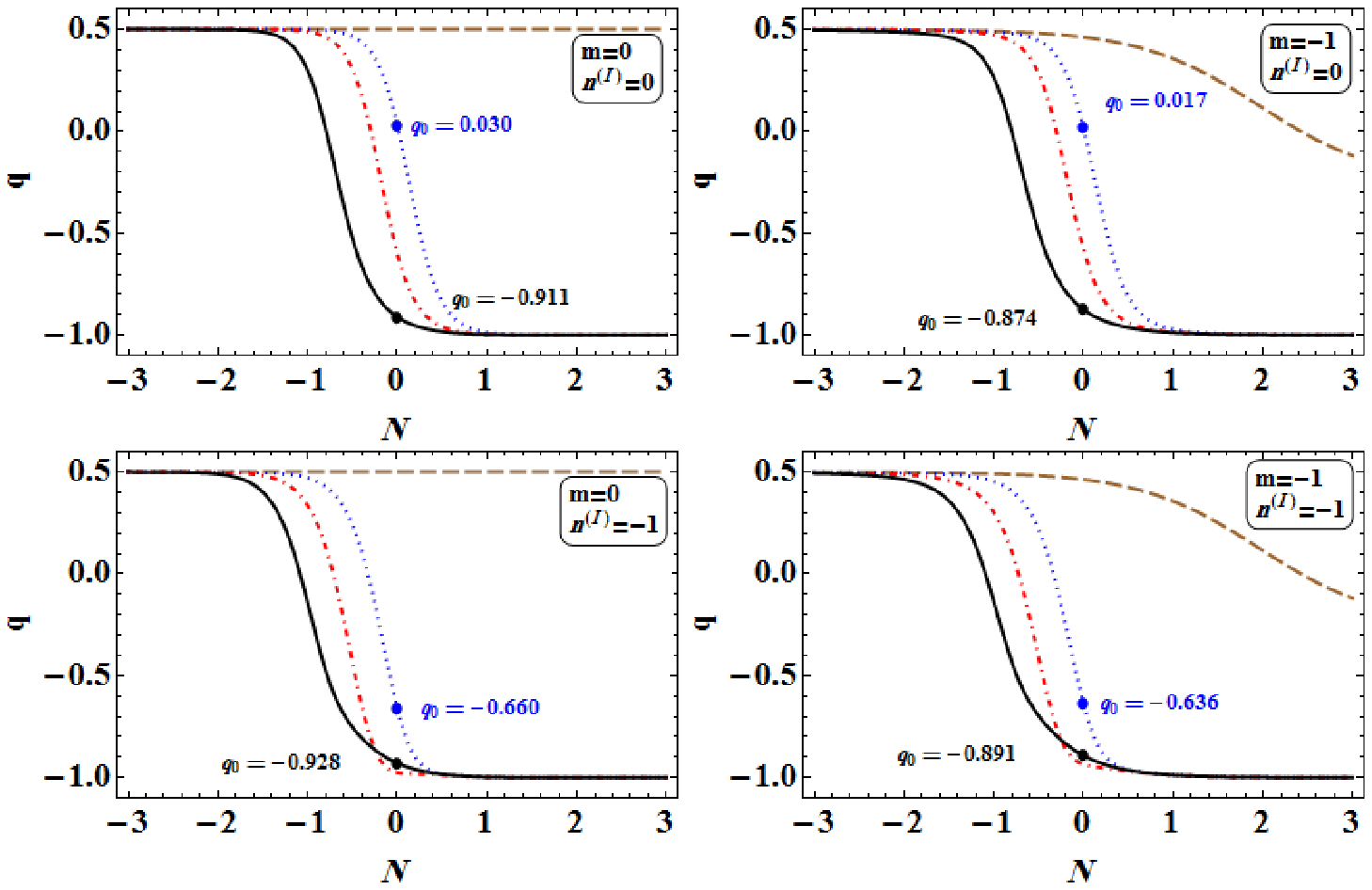,width=12cm}
\caption{(color online). \footnotesize {The evolution of the
deceleration parameter of model I for different negative values of
the parameters $\mathbf{m}$ and $\mathbf{n^{(\textrm{I})}}$.
The curves are plotted for $K=0$, $K=0.33$, $K=0.77$ and $K=0.99$
corresponding to brown dashed line, blue dotted line for,
red dot-dashed line and black solid line. Moreover, present
values indicated by small solid circles. All diagrams are
drawn for the present baryonic matter density
$\Omega_{0}^{\textrm{(b)}}=0.05$. The case $K=0$ corresponds
to the presence of only the baryonic and CDM. In this case,
the deceleration parameter is constant $0.5$ in GR (which corresponds to
$\mathbf{m},\mathbf{n^{(\textrm{I})}}=0$). On the contrary,
decreasing the value of $\mathbf{m}$ leads to a varying deceleration
parameter in $f(R,T)$ gravity. By decreasing both parameters,
we obtain better results for the only cases with $K<0.5$,
however, decreasing only in $\mathbf{n^{(\textrm{I})}}$ improves the
result for all values $0<K<1$.}}
\label{fig1}
\end{figure}

\begin{figure}
\epsfig{figure=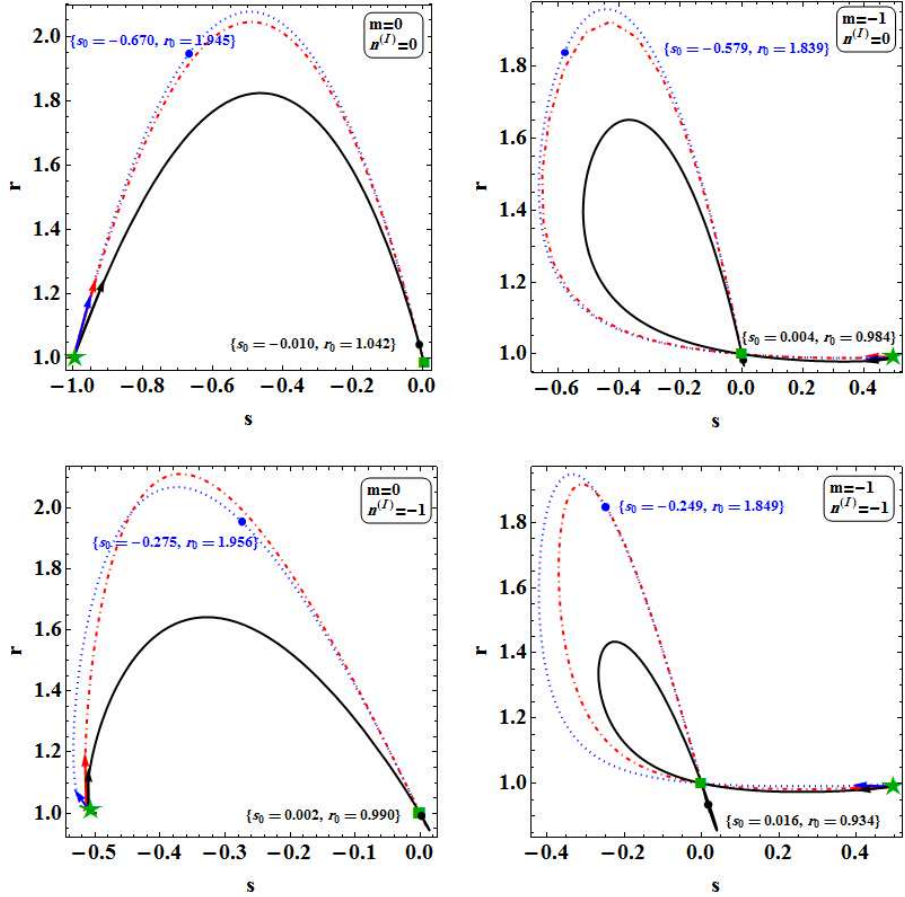,width=12cm}
\caption{(color online). \footnotesize {Statefinder diagnosis of
model I in the $(s,r)$ plane. In this plane the parameter $r$ corresponds to
the vertical axis and the values of parameter $s$ form the horizontal axis. Each
trajectory start from a star symbol and end up at a boxed symbol. the star symbol
denotes the initial value for the trajectories. There is three different cases;
some trajectories start from (-1,1) (upper left panel), some start from
(-0.5,1) (lower left panel) and other trajectories begin from (+0.5,1)
(right diagrams). However, all trajectories terminate at the boxed symbol
place with coordinate $(0,1)$ which belongs to the $\Lambda$CDM model.
The diagram in lower right panel shows that decreasing the values of
$\mathbf{m}$ and $\mathbf{n^{(\textrm{I})}}$ reduces the distance to the
$\Lambda$CDM fixed point which for $K>0.5$ is not very effective.}}
\label{fig2}
\end{figure}

The evolution of the deceleration parameter
(\ref{qI}) is presented in Fig.~\ref{fig1}. An interesting
result is that, the value of the deceleration parameter
for the models with $K=0$ is running when $\mathbf{m}<0$.
The curve corresponding to $K=0$, is sensitive to negative values of $\mathbf{m}$.
Plots show that there is a transition between the value $1/2$ and negative values
of $q$. This transition for $K=0$ is relatively slower than for the other value of $K$;
the larger values the parameter $\mathbf{m}$ gets, the later the transition
occurs. In addition to the mentioned effect of negative values of
$\mathbf{m}$ for curves with $K=0$, the overall effects of the
parameters $\mathbf{m}$ and $\mathbf{n^{(\textrm{I})}}$ can be distinguished. To this
end, the present values of the deceleration function $q_{0}$
in Fig.~\ref{fig1} (and the statefinder parameters $r_{0}$ and $s_{0}$ in
Fig.~\ref{fig2}), are specified for some curves. Comparing upper left corner diagram
with lower left one in Fig.~\ref{fig1}, shows that the net effect of
$\mathbf{n^{(\textrm{I})}}(<0)$ is to decrease the values of
$q_{0}$. Since there is a linear relation between the deceleration and
EoS parameters, there exists a similar mutual behavior for the effect
of $\mathbf{m}$ when $\mathbf{n^{(\textrm{I})}}=0$.

In Fig.~\ref{fig2}, we present the evolution of the statefinder parameters
in the $(s,r)$ plane for $K=0.33, 0.77, 0.99$. In this plane, the values of $s$ are mapped on a
horizontal axis and a vertical axis determines the values of
$r$. In the upper left panel, we plot the diagrams
that hold for GR, in order to see possible deviations. In each panel
besides the present values (which are indicated by small solid
circle), we represent starting point with
a star symbol (corresponding to a point in the early times), the
position of the $\Lambda$CDM model (corresponding to future)
with a solid small box and the direction of the evolution of the trajectories
with colored arrows. The starting point corresponding to the early
times is obtained by the condition $a\rightarrow 0$. In the limit
$a\rightarrow 0$, the statefinder parameter $r$ reads
\begin{align}\label{rI-early}
r^{(\textrm{I})}(a\rightarrow 0)=1+\frac{9\mathbf{m}
\Omega_{0}^{\textrm{(b)}}a^{3/2}}{8\Big{(}\Omega_{0}^{\textrm{(b)}}-
\frac{\sqrt{1-K}(1+(\mathbf{m}-1)\Omega_{0}^{\textrm{(m)}})}
{2\sqrt{6}K^{3/2}\mathbf{n}^{(\textrm{I})}-1}\Big{)}},
\end{align}
whence, for the early times we get $r=1$.
However, the limit of $s$ for $a\rightarrow 0$ depends on the value of
the parameters $\mathbf{m}$ and $\mathbf{n}^{(\textrm{I})}$; for zero
value for both parameters we have $s=-1$, if only the former is vanished $s=-1/2$, and otherwise
we have $s=+1/2$. Therefore, for non-interacting SCG and baryonic
matter, in the background of $f(R,T)$ theory of gravity, the trajectories of
the statefinder plane belongs to three different subclasses; some trajectories
have the initial value $(s=-1,r=1)$, and the other two have $(s=\pm 1/2,r=1)$, as the
initial values. However, all trajectories terminate at the $\Lambda$CDM fixed point,
since the statefinder parameters in the limit $a\rightarrow \infty$ take the
following forms
\begin{align}
&r^{(\textrm{I})}(a\rightarrow \infty)=1+\mathcal{C}_{1}\Big{(}K,\mathbf{m},
\mathbf{n}^{(\textrm{I})},\Omega_{0}^{\textrm{(b)}}\Big{)}a^{-3/2},\label{rI-late}\\
&s^{(\textrm{I})}(a\rightarrow \infty)=\mathcal{C}_{2}\Big{(}K,\mathbf{m},
\mathbf{n}^{(\textrm{I})},\Omega_{0}^{\textrm{(b)}}\Big{)}a^{-3/2},\label{sI-late}
\end{align}
where $\mathcal{C}_{1}$ and $\mathcal{C}_{2}$ are some functions of their
arguments. Here, in model I, there are different scenarios for the evolution of the Universe,
with more or less the same feature in their deceleration parameters,
but completely different feature in the ($r,s$) plane; the Universe begins
with different initial points, however, terminates in the
same point in the late times, mimicking the cosmological constant mode.
Lateral comparison of diagrams in Fig.~\ref{fig2}, shows that for a specific
value of $K$, the net effect of negative values
of $\mathbf{m}$ is an increase in the value of $s_{0}$, and a
decrease in the value of $r_{0}$. A columnar view shows that negative values of
$\mathbf{n^{(\textrm{I})}}$ would increase the value of $s_{0}$. Nevertheless,
in these cases the value of $r_{0}$ grows for $0.17<K<0.33$
(the maximum value $r_{0,max}\simeq 1.76$
occurs in $K\simeq 0.24$ ), and decreases otherwise. Diagonal comparison
demonstrates that the distance to the $\Lambda$CDM fixed point is shorter
when both parameters $\mathbf{m}$ and $\mathbf{n}^{(\textrm{I})}$ are
turned on. Such an effect is not seen for large $K$; negative values of
$\mathbf{m}$ and $\mathbf{n}^{(\textrm{I})}$ makes the
distance of the model to the $\Lambda$CDM fixed point slightly larger comparing
with GR. However, the models with $K\rightarrow 1$ have shorter distance.

\subsection{model $f^{\textrm{(II)}}\Big{(}R, T^{\textrm{(b, G)}}\Big{)}$}\label{Model II}
\begin{figure}[h]
\epsfig{figure=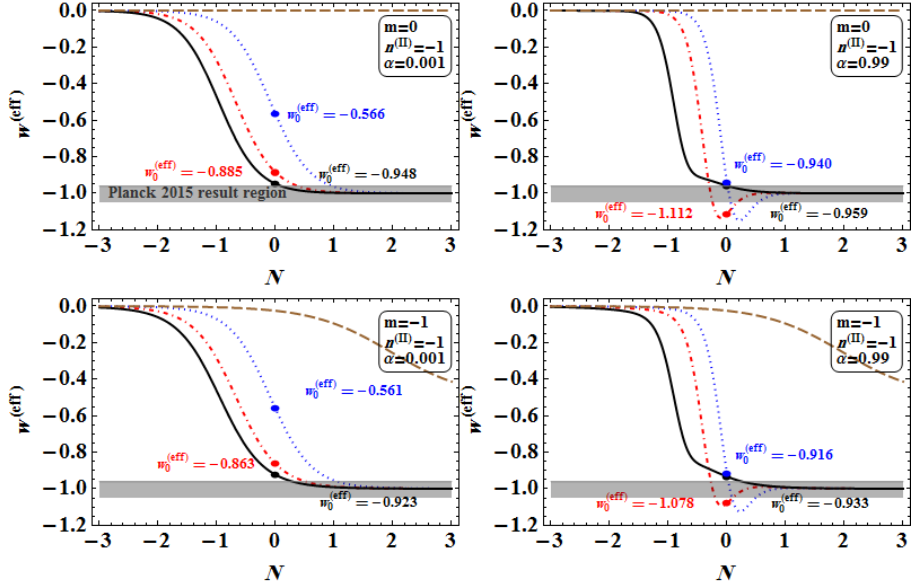,width=12cm}
\caption{(color online). \footnotesize {The evolution of the
effective EoS parameter for model II which uses the
non-interacting baryonic matter together with GCG in the
high pressure regime. The curves are plotted for $K=0$
(brown dashed line), $K=0.33$ (blue dotted line), $K=0.77$
(red dot-dashed line) and $K=0.99$ (black solid line),
for $\mathbf{m}=0,-1$, $\mathbf{n^{(\textrm{II})}}=-1$
and $\alpha=0.001,0.99$. The present values are also
depicted with the corresponding colors. The valid range of values reported by Planck 2015
measurements, ($-1.051\leq w_{0, \textrm{Planck}}^{(\textrm{eff})}\leq-0.961$) is
indicated by the gray region. An increase in
the value of $\alpha$, leads to a decrease in the value of
$w_{0}^{\textrm{(eff)}}$.}}
\label{wii}
\end{figure}

\begin{figure}[h]
\epsfig{figure=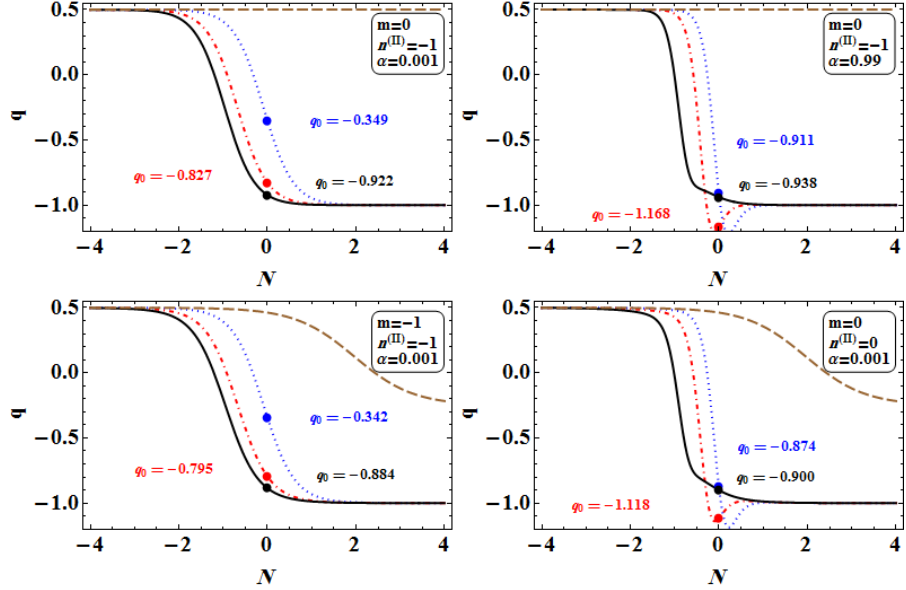,width=12cm}
\caption{(color online). \footnotesize {The evolution of the
deceleration parameter for type II models for different values of
the parameters $K$, $\mathbf{m}=0,-1$ and $\mathbf{n^{(\textrm{II})}}=-1$.
The curves are plotted for $K=0$, $K=0.33$ $K=0.77$ and $K=0.99$,
corresponding to brown dashed line, blue dotted line for,
red dot-dashed line and black solid line. In this case,
a lateral view shows that the value of the deceleration
parameter decreases when parameter $\alpha$ increases. And,
a vertical view shows that decreasing $\mathbf{m}$, results in increasing
the value of $q_{0}$. }}
\label{fig3}
\end{figure}

In this case besides the model coupling constants $\mathbf{m},\mathbf{n^{(\textrm{II})}}$ and
parameter $K$, there is another parameter, $\alpha$, as the forth constant. Therefore, the net
effects of this parameter should be considered. The related figures of the effective EoS parameter
$w^{(\textrm{eff})}$ for this model are illustrated in Fig.~\ref{wii}. Again, the curves
for $K=0,~0.33,~K=0.77$ and $K=0.99$, corresponding to brown dashed, blue dotted, red dot-dashed
and black solid lines are plotted, respectively. Here, the diagrams for $\alpha=0.001,0.99$
are drawn. It is clear from equations (\ref{qI}), (\ref{qII}) and (\ref{qIII}), that the deceleration and
accordingly the effective EoS parameters are the same for models I,
II and III, when $\mathbf{n}=0$'s go to zero. Hence, in this case
the diagrams for $w^{\textrm{(eff)}}$ and $q$ are not redrawn in model(s)
II (and III in the later subsection). We arrange the diagrams the same
as model I, however in this case the parameter $\alpha$
increases horizontally. An increase in the value of $\alpha$ leads to
a decrease in the value of $w_{0}^{\textrm{(eff)}}$, and this is clearly evident
for the relatively smaller values of $K$. In this model we obtain
more acceptable results in all range $0<\alpha<1$; upper left diagram shows
that even for $\alpha=0.001$ we have $w_{0}^{\textrm{(eff)}}\simeq-0.9$
(red dot-dashed line), the result that met in model I only for $K\rightarrow1$.
By comparing the effective EoS diagrams between model I and II we see that
the value of $w_{0}^{\textrm{(eff)}}$ decreases when the value of $\alpha$ increases,
and this means that we can construct more acceptable models for larger values of
$\alpha$.
In Fig.~\ref{fig3}, the related diagrams of
the deceleration parameter $q=q(a;\alpha,K,\mathbf{m},\mathbf{n^{(\textrm{II})}})$, for
$\mathbf{m}=0,-1$, $\mathbf{n^{(\textrm{II})}}=-1$ and $\alpha,K>0$ are drawn. In the lateral view, the
diagrams for the same values of parameters $\mathbf{m},\mathbf{n^{(\textrm{II})}}$ for two values
$\alpha=0.01,0.99$ are pictured. In the columnar view, the plots are arranged for the same value of
$\alpha$ and different values of $\mathbf{m}$. The present value of the
deceleration parameter $q_{0}$ is displayed in each plot, which can be calculated as
\begin{align}\label{q0-I1}
q_{0}^{(\textrm{II})}=-1+\frac{3}{4}\left[\Omega _{0}^{\textrm{(b)}}(2-\mathbf{m})+\frac{2(1-K)\Big{(}1+
\Omega _{0}^{\textrm{(b)}}(\mathbf{m}-1)\Big{)}\Big{(}1+\sqrt{27}K^{3/2}
\alpha\mathbf{n^{(\textrm{II})}}\Big{)}}{1-\sqrt{12} K^{3/2} \mathbf{n^{(\textrm{II})}}}\right],~~~~~~N=0,
\end{align}
and for the special case $\mathbf{n^{(\textrm{II})}}=0$ it can be rewritten as
\begin{align}\label{q0-I2}
q_{0}^{(\textrm{II})}=\frac{3}{4}\Omega _{0}^{\textrm{(b)}}\left(1-2K\right)\mathbf{m}+
\frac{1}{2} \left(1-3K(1+\Omega _{0}^{\textrm{(b)}})\right),~~~~~~N=0,~~~~\mathbf{n^{(\textrm{II})}}=0.
\end{align}
It is obvious that the value of $q_{0}$ does not depend on $\alpha$ for the
models with $\mathbf{n^{(\textrm{II})}}=0$, (that is we have
$q_{0}^{(\textrm{I})}=q_{0}^{(\textrm{II})}=q_{0}^{(\textrm{III})}$
for $\mathbf{n}$'s$=0$). The related plots for $\mathbf{n^{(\textrm{II})}}\neq0$ are presented
in the first row of Fig.~\ref{fig3}. As the plots show, for the cases with
$\mathbf{n^{(\textrm{II})}}\neq0$ an increase in the value of $\alpha$ results
in a decrease in the value of $q_{0}$. The models with observationally
inconsistent value of $q_{0}$ can be put in an acceptable regime
(for which we have $q_{0}\simeq -0.6$)
by choosing a value for parameter $\mathbf{n^{(\textrm{II})}}$, properly.
This arbitrariness can not be met in GR regime, where $f(R,T)=R$.
In GR, observationally consistent values can be achieved only for
$K\simeq 1$, irrespective of the value of $\alpha$. However in $f(R,T)$ gravity
one can attain the consistent Chaplygin gas models with arbitrary value of $K$ only
by a suitable choice of the coupling constant $\mathbf{n^{(\textrm{II})}}$, (compare
the diagrams in the top panels in Fig.~\ref{fig1} with the diagrams in Fig.~\ref{fig3}).
Comparing the first row with the second one in Fig.~\ref{fig3}, shows that
turning parameter $\mathbf{m}$ on increases the values of the
deceleration parameter. Equation (\ref{q0-I1}) shows that the present
value of the deceleration (and also the effective EoS) parameter
for $\mathbf{n^{(\textrm{II})}}=-1$ is not as simple as the result
given in (\ref{q0-I2}), and therefore the effect of variation of $\mathbf{m}$
is not obvious for each value of $K$. However, as diagrams in
Fig.~\ref{fig3} show, decreasing the value of $\mathbf{m}$ increases the value
of $q_{0}$. Comparing the diagrams for model I in Fig.~\ref{fig1} with
the diagrams in Fig.~\ref{fig3}, one can find more acceptable results in model II for
large values of $\alpha$.

\begin{figure}
\epsfig{figure=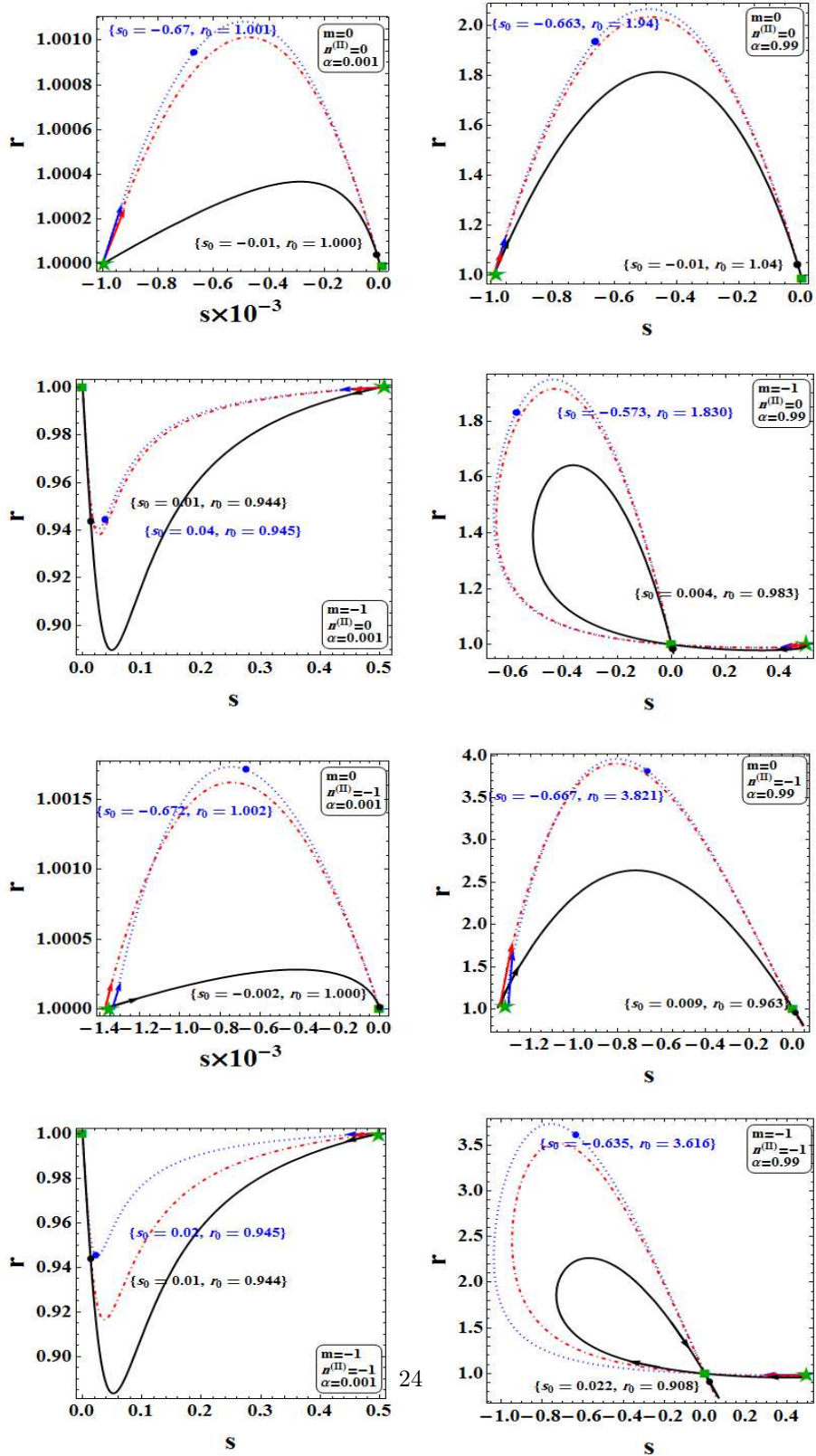,width=12cm,height=21.5cm}
\caption{(color online). \footnotesize {Cosmological trajectories of
model II in the $(s,r)$ plane. The plots show that an
increase in the value of $\alpha$, has a significant effect
on the values of statefinder parameters for smaller values of $K$, and
accordingly, these models have a distinct situation with respect to
the $\Lambda$CDM model.}}
\label{fig6}
\end{figure}

In Fig~\ref{fig6} we have depicted ($r,s$) plane diagrams for different values $K=0.33,0.77,0.99$,
$\mathbf{m},\mathbf{n^{(\textrm{II})}}=0,-1$ and $\alpha=0.001, 0.99$.
These diagrams have been arranged in such a way that the parameter $\alpha$
varies in the horizontal view and the coupling constants $\mathbf{m},\mathbf{n^{(\textrm{II})}}$
in the vertical view. Comparison of the lateral diagrams shows that in the high pressure regime
a growth in the value of $\alpha$ would result in a decrease in the value of $s_{0}$ and increases the value of
$r_{0}$ for $\mathbf{n^{(\textrm{II})}}=0$. However, when $\mathbf{n^{(\textrm{II})}}<0$ this
feature cannot be seen for large values of $K$. This means that the distance to
the $\Lambda \rm{CDM}$ fixed point becomes so long when $\alpha$ approaches to $1$ and $k$ gets away from $1$.
That is, the models with large values of $\alpha$ and small values of $K$ are distinguishable
from the $\Lambda \rm{CDM}$ model in the high pressure regime. However, since the $\Lambda \rm{CDM}$ is the mostly
accepted cosmological model, these models cannot be very interesting. The upper left panel holds for
GR regime which resembles the $\Lambda \rm{CDM}$ model, since, we have $s_{0}\propto -10^{-3}$ and
$r_{0}\approx 1$. In the vertical view, we see that decreasing the values of $\mathbf{m}$ does not lead to
a significant variations of the parameters $r_{0}$ and $s_{0}$. Therefore, decreasing in the
value of $\mathbf{m}$ does not affect the distance of underlying model. Furthermore,
when $\mathbf{n^{(\textrm{II})}}$ decreases, the present
fixed point $(s_{0},r_{0})$ gets farther from the $\Lambda \rm{CDM}$ one only for large values of $\alpha$ and
small values of $K$.
\subsection{model $f^{\textrm{(III)}}\Big{(}R, T^{\textrm{(b, G)}}\Big{)}$}\label{Model III}
\begin{figure}[h]
\epsfig{figure=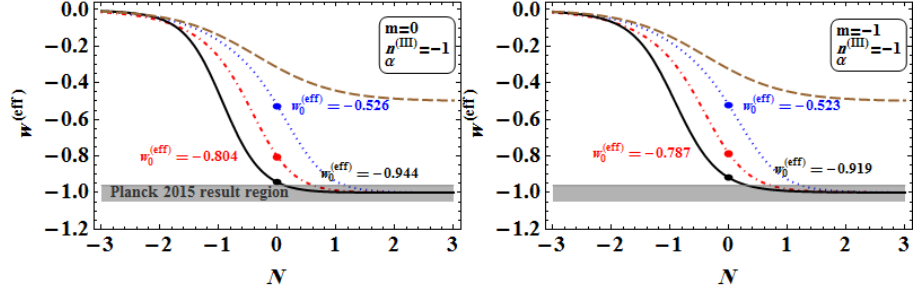,width=12cm}
\caption{(color online). The evolution of the
EoS parameter for model III. Here the values of $\alpha$ has no effect on the
values of $w^{\textrm{(eff)}}$.}
\label{fig7}
\end{figure}

\begin{figure}[h]
\epsfig{figure=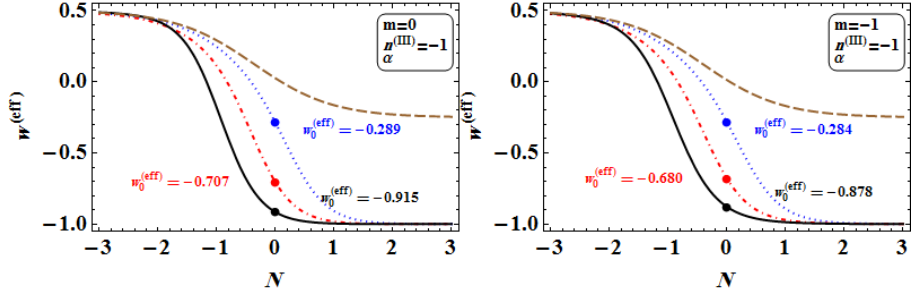,width=12cm}
\caption{(color online). The evolution of the
deceleration parameter for model III. The values of $q_{0}$ are independent of the
values of $\alpha.$}
\label{fig8}
\end{figure}

\begin{figure}
\epsfig{figure=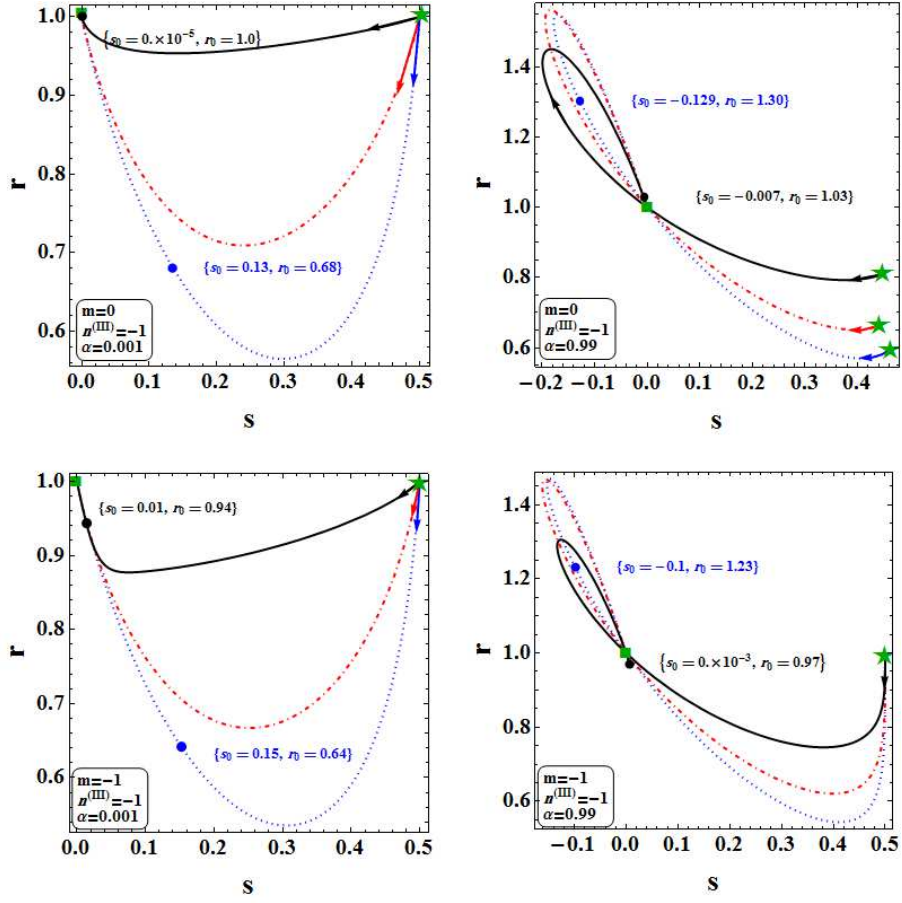,width=12cm}
\caption{(color online). \footnotesize {Statefinder diagnosis of
model III in the $(s,r)$ plane. Only the diagrams for
$\mathbf{n^{(\textrm{III})}}=-1$ are plotted since the case
$\mathbf{n^{(\textrm{III})}}=0$ is the same as the results of
model II when $\mathbf{n^{(\textrm{II})}}=0$. Horizontal view
shows that if the value of $\alpha$ increases, the distance to the
$\Lambda$CDM fixed point will slightly enlarge. Also, a decrease in the value
of $\mathbf{m}$ has no effective impact on the present values.}}
\label{fig9}
\end{figure}

This case differs from model II only when $\mathbf{n^{(\textrm{III})}}\neq0$. Both
models II and III have the same behavior for $\mathbf{n^{(\textrm{II})}}=
\mathbf{n^{(\textrm{III})}}=0$. The other important point about model III is that,
the present values of the deceleration parameter $q_{0}$ and therefore $w_{0}$,
are independent of the values of $\alpha$ for arbitrary values of $\mathbf{n^{(III)}}$.
The present value of the deceleration parameter for this model reads
\begin{align}\label{q0-III}
&q^{(\textrm{III})}_{0}=\frac{3}{4 (1-2 \mathbf{n^{(\textrm{III})}})} \Bigg{[}\Omega _{0}^{\textrm{(b)}} \left(1-2 K (1-\mathbf{n^{(\textrm{III})}})\right)\mathbf{m}-2 \left(K(\Omega _{0}^{\textrm{(b)}}-1)+2 \Omega _{0}^{\textrm{(b)}}+1\right)\mathbf{n^{(\textrm{III})}}+\Bigg.\nonumber\\
&\Bigg.2 \left(K(\Omega _{0}^{\textrm{(b)}}-1)+1\right)\Bigg{]},~~~~~~N=0,
\end{align}
whence we observe that the parameter $\alpha$ does not play any
role in the value of $q^{(\textrm{III})}_{0}$. In Fig~\ref{fig7}
and Fig~\ref{fig8} we present the related diagrams of the effective
EoS and the deceleration parameters for $\mathbf{n^{(\textrm{III})}}=-1$,
$\mathbf{m}=0,-1$, respectively. By indicating the parameter $\alpha$ on
diagrams, we emphasize that the present values do not depend on $\alpha$.
when parameter $\mathbf{m}$ is switched on,
an increase in the values of $w_{0}$ and $q_{0}$ would happen.
Such an effect appears in model II, as well. Since, when
$\mathbf{n}$'s$=0$, we have the same values of
$w_{0}$ and $q_{0}$ for the three models, therefore to understand the effect of
decreasing the value of $\mathbf{n^{(\textrm{III})}}$ from zero
to an arbitrary negative value (here $-1$), one can compare
the diagrams of Fig.~\ref{fig8} (or Fig.~\ref{fig7} for the EoS parameter
of model III) with the corresponding ones in Fig.~\ref{fig1} (or Fig.~\ref{wi}
for the EoS parameter of model I). We see that decreasing the values of
$\mathbf{n^{(\textrm{III})}}$ would reduce the values of $q_{0}$ and $w_{0}$,
and the net effect is more significant for smaller values of $K$.
Thus, again decreasing $\mathbf{n^{(\textrm{III})}}$ to negative values
can help to construct more reliable models with admissible values of
$w_{0}$ and $q_{0}$.
In Figure~\ref{fig9} we present the related
diagrams for the statefinder parameters of model III. These diagrams are plotted for $\mathbf{n^{(\textrm{III})}}=-1$,
since the diagrams for $\mathbf{n^{(\textrm{III})}}=0$ are the same as the diagrams for model II with
$\mathbf{n^{(\textrm{II})}}=0$. Comparing the plots of Fig.~\ref{fig9} with four upper diagram
in Fig.~\ref{fig6} indicates that a decrease in the value of $\mathbf{n^{(\textrm{III})}}$ makes the
distance of model III to the $\Lambda$CDM model shorter for large values of $\alpha$ and longer
for small values of this parameter. As diagrams of Fig.~\ref{fig9} show, an increase in the value of $\alpha$, decreases the
value of $s_{0}$ and makes $r_{0}$ to get larger values. The overal behavior of this, is to lengthen the distance to
the $\Lambda$CDM fixed point. Nevertheless, the effect is weak for larger values of
$K$. A vertical view indicates that decreasing the parameter $\mathbf{m}$, has no remarkable effect on
the present values of the statefinder parameters, though it leads the distance to become slightly
shorter for larger values of $\alpha$ and smaller value of $K$.
\section{Luminosity distance and Hubble parameter tests}\label{test}
So far we have discussed the $f(R,T)$ cosmological solutions when a generalized Chaplygin
gas along with a baryonic matter are present as the whole matter content. We have obtained
three different forms for $f(R,T)$ function based on the EMT conservation and
subsequently the cosmological consequences of these models have been extracted
from a theoretical point of view. In this section, these solutions will be  compared to
SNIa cosmological data and the compilation of Hubble parameter measurements.
Supernovae type Ia as accurate standard candles are the most direct probe to detect signatures
of the accelerated expansion of the Universe as well as DE. Therefore, Dark energy scenarios in the first step can be examined
by comparing their predicted luminosity distance to the observational data. In this regard, we apply the
chi-square test of the goodness of fit of the Hubble parameter and modulus distance quantity defined by equations
(\ref{EI})-(\ref{EIII}) for models I, II and III to the observational data. The procedure of fitting
can be summarized as follows: The supernova observations include $N$ apparent
magnitude $m_{i}$ which can be translated to a quantity which is called
``distance modulus $\mu_{i}$", the corresponding redshift $z_{i}$ and their errors. Theoretically, these quantities
satisfy the following equation
\begin{align}\label{m-mu}
\mu(z)\equiv m(z)-M=25+5\log\left(\frac{d_{L}(z)}{1 Mpc}\right),
\end{align}
where the luminosity distance $d_{L}$ is defined as
\begin{align}\label{lumin}
d_{\textrm{L}}(z)=\frac{c}{H_{0}}(1+z)\int_{0}^{z}\frac{dz'}{H(z')}.
\end{align}
Generally, for a given model, the Hubble parameter $H(z)$ can be written as a function of some parameters
like those presented in our cases. For example, we have $H(z; \mathbf{m}, \mathbf{n}'s,K,\alpha)$ in this paper. The idea is
to find the best fit values of the model parameters which lead to the most consistency between
the predicted curves of the modulus distance as well as the Hubble parameter (by the virtue of the underlying theoretical model) and the observational
data. To this aim, we minimize the quantities
\begin{align}\label{chi-mu}
\chi^{2}_{\textrm{$\mu$}}(\mathbf{m}, \mathbf{n}'s,K,\alpha)=\sum_{i=1}^{N}\frac{\left[\mu^{(\textrm{o})}(z_{i})-
\mu^{(\textrm{th})}(z_{i})\right]^{2}}{{\sigma_{\textrm{H,i}}^{2}-\sigma_{\textrm{mz,i}}^{2}}},
\end{align}
and
\begin{align}\label{chi-Hubble}
\chi_{\textrm{H}}^{2}(\mathbf{m}, \mathbf{n}'s,K,\alpha)=\sum_{j=1}^{N'}\frac{\left[H^{(\textrm{o})}(z_{j})-
H^{(\textrm{th})}(z_{j})\right]^{2}}{{\sigma_{\textrm{$\mu$},j}^{2}}},
\end{align}
where $\mu^{(\textrm{o})}(z_{i})$ and $H^{(\textrm{o})}(z_{i})$ are obtained from observations,
$\mu^{(\textrm{th})}(z_{i})=\mu^{(\textrm{th})}(z_{i};\mathbf{m}, \mathbf{n}'s,K,\alpha)$
and $H^{(\textrm{th})}(z_{i})=H^{(\textrm{th})}(z_{i};\mathbf{m}, \mathbf{n}'s,K,\alpha)$ denotes the distance modulus and
the Hubble parameter predicted by theory, $z_{i}$ is the redshift of an event obtained from observation,
$\sigma_{\textrm{$\mu$},i}$ and $\sigma_{\textrm{H},i}$ are the errors caused by measurement and
$\sigma_{\textrm{mz,i}}$ is dispersion in the distance modulus.
There are some packages on the web that can help one to minimize the chi-square function for the
best fit parameters of theory. For example, in~\cite{pack1} one can find the self-study package
``CoChiSquare alpha" which is written in both Mathematica~\cite{Math} and Python~\cite{pyth} languages, or the one
which is released in~\cite{pack2}. After minimizing the chi-square parameter for the best fit values, if
$\chi^{\textrm{2}}_{\textrm{min}}(\mathbf{\overline{m}}, \overline{\mathbf{n}}'s,\overline{K},\overline{\alpha})/d.o.f\lesssim1$
then we have a good fitting and the theory is consistent with the data. ``d.o.f" stands for the number of degrees of
freedom which is equal to $N-n$ in which $N$ is the total number of measurements, $n$ is the number of parameters of theory, and
the barred parameters in the argument are the best fit values.\\

\begin{figure}[h]
\centering
\epsfig{figure=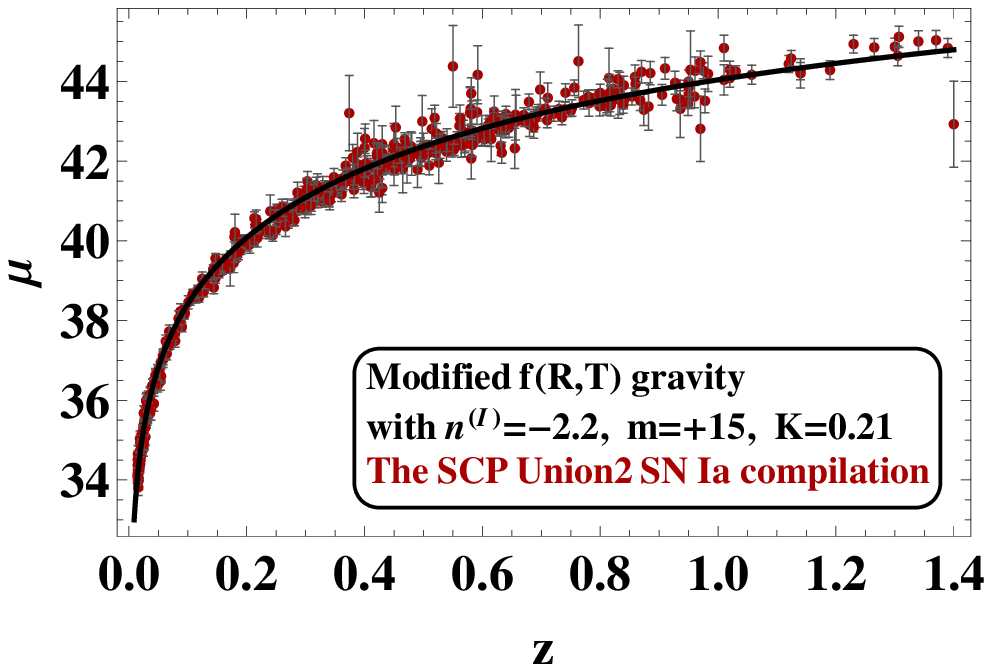,width=5.8cm}\hspace{4mm}
\epsfig{figure=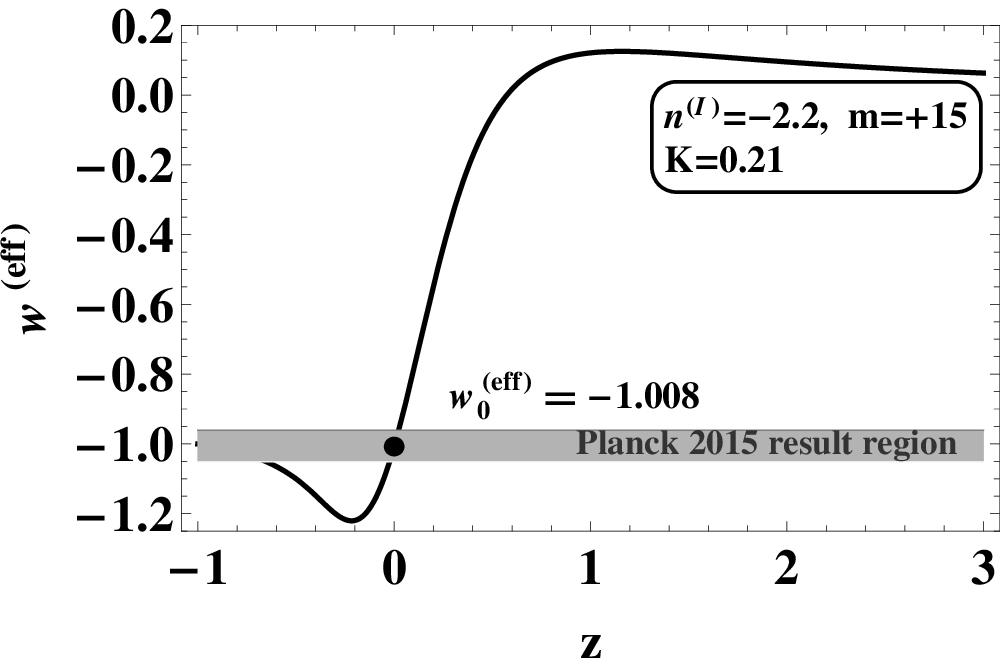,width=5.8cm}\vspace{2mm}
\epsfig{figure=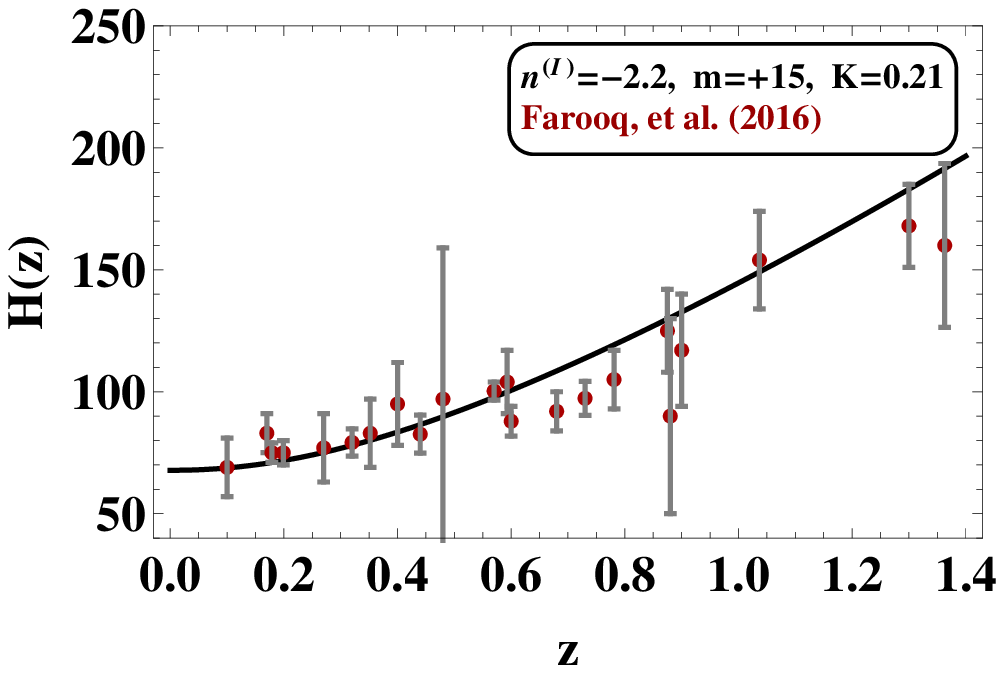,width=5.8cm}
\caption{(color online). \footnotesize {Cosmological diagrams for the best fit
values of model I parameters obtained from the modulus distance chi-square
test. The upper left panel shows the modulus distance and the Union 2 compilation data
which are shown in red points. The upper right panel presents the effective EoS
parameter. In this plot the valid range of values reported by Planck 2015
measurements ($-1.051\leq w_{0, \textrm{Planck}}^{(\textrm{eff})}\leq-0.961$) is
indicated by the gray region. The lower one depicts the Hubble parameter
along with the Hubble cosmological data which is illustrated in red color.}}
\label{fig10}
\end{figure}

\begin{figure}[h]
\centering
\epsfig{figure=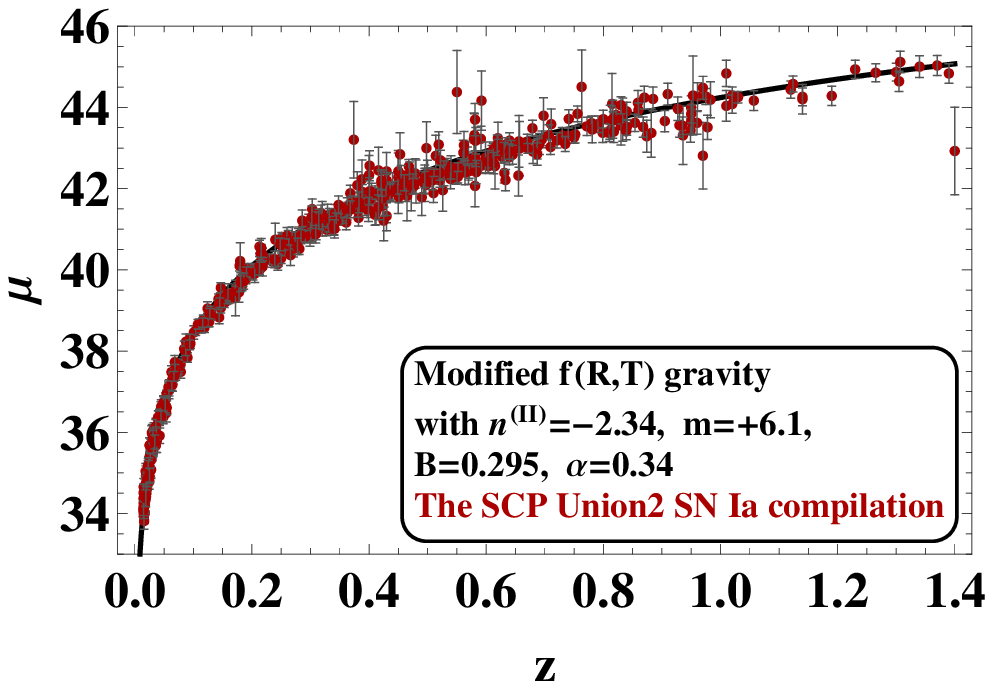,width=5.8cm}\hspace{4mm}
\epsfig{figure=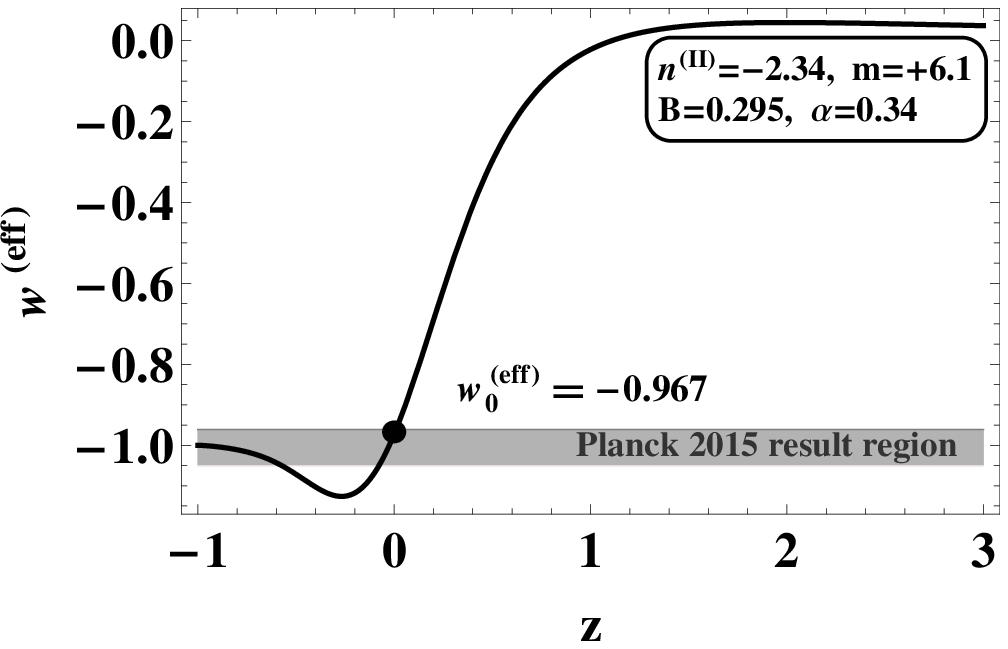,width=5.8cm}\vspace{2mm}
\epsfig{figure=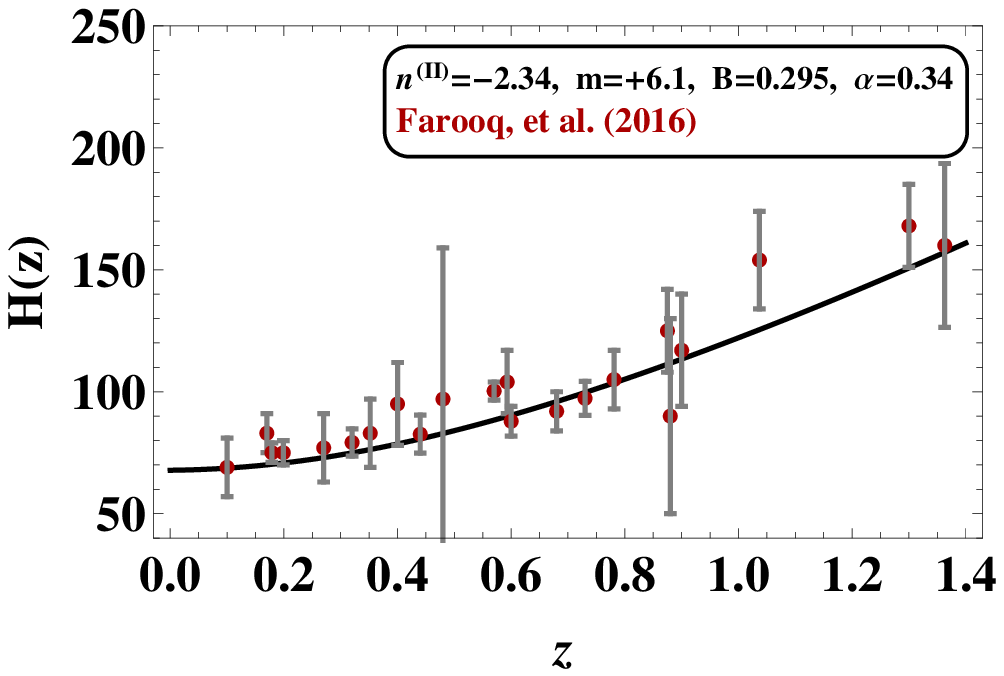,width=5.8cm}
\caption{(color online). \footnotesize {Cosmological diagrams for the best
fit values of model II parameters obtained from the modulus distance chi-square
test.}}
\label{fig11}
\end{figure}

In our study we have used the Union 2 sample which consists of 557 type Ia supernova data~\cite{union} as well as
28 Hubble parameter measurements\footnote{The original data can be found in Refs.\cite{H1,H2,H3,H4,H5,H6,H7,H8}} which are
collected in Ref.~\cite{farooq}. We minimize the chi-square function to find the best fit values of
$\mathbf{m}$, $\mathbf{n}$'s, $K$ and $\alpha$ parameters for the fixed values
$H_{0}=67.8$ $km s^{-1}Mpc^{-1}$, $\Omega_{0}^{(\textrm{b})}=0.05$ and $c=2.99\times10^8 ms^{-1}$.
The results are summarized in Table~\ref{Tab1}. In this table in addition to
$\chi^{\textrm{2}}_{\textrm{($\mu$ ,H) min}}$ and $\chi^{\textrm{2}}_{\textrm{($\mu$, H) min}}/d.o.f$
the effective EoS and the value of deceleration parameter for the best fit values of model parameters
(which are obtained in the modulus distance and Hubble chi-square tests) are shown.
\begin{sidewaystable}
\begin{center}
\caption{Results of the modulus distance and Hubble chi-square tests for models I, II and III.}
\begin{tabular}{l @{\hskip 0.05in} c@{\hskip 0.1in} c@{\hskip 0.15in} l@{\hskip 0.15in} c@{\hskip 0.15in} c@{\hskip 0.15in} c@{\hskip 0.15in} c}\hline\hline

Models&    $\chi^{\textrm{2}}_{\textrm{($\mu$,H) min}}$&   $ \chi^{\textrm{2}}_{\textrm{($\mu$,H) min}}/d.o.f$&
Best fit values of the model parameters&    $q_{0}$&    $w^{(\textrm{eff})}_{0}$&$s_{0}$&$r_{0}$\\[0.5 ex]
\hline
Model I&20.667&0.827&$K_{\textrm{H}}=0.180, m_{\textrm{H}}=-7.68, n_{\textrm{H}}^{(\textrm{I)}}=-4.66,
\alpha----$&-0.604&-0.736&-0.12&1.4\\[0.75 ex]
Model I&558.024&1.0073&$K_{\textrm{$\mu$}}=0.210, m_{\textrm{$\mu$}}=15, n^{(\textrm{I})}_{\textrm{$\mu$}}=-2.2,
\alpha----$&-1.01&-1.008&-0.65&3.97\\[0.75 ex]
Model II&23.680&0.947&$K_{\textrm{H}}=0.35, m_{\textrm{H}}=-7.10, n_{\textrm{H}}^{(\textrm{II})}=-2.56,
\alpha_{\textrm{H}}=-0.33$&-0.267&-0.511&0.31&0.28\\[0.75 ex]
Model II&558.774&1.0104&$K_{\textrm{$\mu$}}=0.295, m_{\textrm{$\mu$}}=6.1, n_{\textrm{$\mu$}}^{(\textrm{II})}=-2.34,
\alpha_{\textrm{$\mu$}}=0.34$&-0.978&-0.967&-0.37&2.62\\[0.75 ex]
Model III&$552.612$&0.9993&$K_{\textrm{$\mu$}}=0.280, m_{\textrm{$\mu$}}=1.0, n_{\textrm{$\mu$}}^{(\textrm{III})}=-1.5,
\alpha_{\textrm{$\mu$}}=0.28$&-0.552&-0.288&0.09&0.77\\[0.75 ex]
\hline\hline
\end{tabular}
\label{Tab1}
\end{center}
\end{sidewaystable}
As shown in Table~\ref{Tab1}, these best fit values for modulus distance test would result in
a good consequence both for models I and II. We have obtained $\chi^{2}_{\textrm{$\mu$}, min}/d.o.f\simeq1.0073$ for model I and
$\chi^{2}_{\textrm{$\mu$}, min}/d.o.f\simeq1.0104$ for model II, which show a relatively good consistency of these models with
the SNIa observations. In these cases, the best fit values give an acceptable present value for the effective EoS.
Nevertheless, the best fit values in the Hubble chi-square tests do not lead to a satisfactory result for $w^{(\textrm{eff})}$
in both cases. The diagrams of the modulus distance along with the Union 2 sample data, the effective EoS and
the Hubble parameter along with the related observational
data~\cite{farooq} for models I and II are drawn in Fig.~\ref{fig10} and Fig.~\ref{fig11}, respectively.
These figures are drawn for $K_{\textrm{$\mu$}}=0.210$, $m_{\textrm{$\mu$}}=15$
and $n_{\textrm{$\mu$}}^{(I)}=-2.2$ in model I and
$K_{\textrm{$\mu$}}=0.295, m_{\textrm{$\mu$}}=6.1, n_{\textrm{$\mu$}}^{(II)}=-2.34, \alpha_{\textrm{$\mu$}}=0.34$ in model II
which minimize the modulus distance chi-square parameter. As is seen, they show a relatively
consistent behavior, at least in low redshifts.

Table~\ref{Tab1} also gives the present values of the statefinder parameters $s_{0}$ and
$r_{0}$ which are calculated for the best fit values of parameters for models I, II and III.
In comparison to the results of the gold sample supernova type
Ia which gives $1.65<r^{\textrm{(Gold)}}_{0}<3.97$ and $-0.86<s^{\textrm{(Gold)}}_{0}<-0.13$~\cite{riess},
the SNLS supernova type Ia data set which results in $0.11<r^{\textrm{(SNLS)}}_{0}<2.69$ and
$-0.61<s^{\textrm{(SNLS)}}_{0}<0.32$~\cite{astier} and
X-ray galaxy clusters analysis with $-1.49<r^{\textrm{(X-ray)}}_{0}<3.06$ and
$-0.56<s^{\textrm{(X-ray)}}_{0}<-0.094$~\cite{rapetti} , we see that model II, namely, $f(R,T)$ gravity with
GCG in high pressure regime shows a good consistency with these data. Only, model II (for the best fit values of parameters
which minimize the modulus distance chi-square) is approximately consistent with all these three data sets.

Similar considerations can be done for model III in search of the observationally acceptable model parameters.
Nevertheless, we will report the results of a full statistical discussion on these models, elsewhere.
The most performed cosmological tests in the literature consist of the
luminosity distance tests, experiments on cosmic microwave background radiation (CMB), the baryon acoustic oscillation (BAO) probes,
investigations about look-back time and the age of the Universe and the growth rate of matter density perturbations.

We conclude this section by commenting that, the model of the Chaplygin gas in the framework of $f(R,T)$ gravity can deserve further investigations.
Some research works have treated this issue~\cite{rul1,rul2,rul3,rul4,rul5,rul6}, where it is shown that the standard Chaplygin gas in the background of GR
may not be consistent with observations,
however, here we have demonstrated that this type of matter can be survived, still in $f(R,T)$ modified theory of gravity. Although, using more precise
statistical techniques for different cosmological tests can give better results.
\section{conclusions}\label{conclusions}
In the present work we have investigated cosmological behavior of the generalized
Chaplygin gas (GCG) in $f(R,T)$ theory of gravity endowed with a homogeneous and
isotropic FLRW spacetime, by means of statefinder diagnosis. More precisely, the
baryonic matter is also included which does not directly interact with GCG. However,
these two fluids interact with each other via a non-standard geometric rule which
is imposed by $f(R,T)$ gravity Lagrangian. Here, the coupling of the trace of whole
matter to geometry would relate the evolution of these two type of matters to each other.

In $f(R,T)$ gravity conservation of energy-momentum tensor (EMT) enforces us
to use a constraint equation which must hold by $f(R,T)$ function. This equation
gives the functionality of $R$ and $T$ by the virtue of which the conservation of
EMT is guaranteed. Only models of type $f(R,T)=g(R)+h(T)$ respect the EMT conservation.
However, we have worked on the special case $f(R,T)=R+h(T)$, although models $f(R,T)=g(R)+h(T)$
are worthy of study as well. Adding a matter trace dependent term to the Ricci scalar
can be accounted for as a correction to the Einstein-Hilbert action.
We solved the constraint equation for two class of models; the models that
employ the standard Chaplygin gas (SCG) and those that make use of GCG. All
related equations (specially the function $f(R,T)$) of the former are exactly
obtained. Nevertheless, the constraint equation can not be exactly solved for the
latter because of the appearance of constant $\alpha$ in the GCG sector. In fact,
for each value of this constant (which may be whether integer or not) there may be
numerous solutions so that dealing with them is time consuming. To avoid such
mathematical problems we approximate the constraint equation in two extreme
limits, the high pressure ($p\gg \rho$) and the high density regimes ($\rho\gg p$).
Of course, we have supposed that under these two extreme situations, the Chaplygin
gas preserves its nature. As a result, we have mainly worked on three classes of Chaplygin gas models;
the models in which SCG is included and we classified them as model I, the GCG models
in the high pressure regime, as models II and finally the GCG models in the high
density regime as models III.

After obtaining and then normalizing the Hubble parameter for each model, we obtained
the deceleration and the statefinder parameters, from which one can differentiate these
models from DE point of view. The statefinder diagnosis is a powerful tool to understand
the behavior of different DE scenarios. In our models, there are some parameters that
enable us to choose the best values for them in order to make the models consistent with
the observational results. Model I contains parameter $\mathbf{m}$ which is related
to the coefficient of the baryonic term in the $f(R,T)$ function, $\mathbf{n^{(I)}}$
which corresponds to the SCG term in the $f(R,T)$ function and $K$ which is included
in the SCG density. In addition to parameters $\mathbf{n^{(II)}}$ and $\mathbf{n^{(III)}}$
which correspond to the GCG terms in the $f(R,T)$ functions and constant $K$ that comes
from GCG density, we have constant $\alpha$ for the two other models. In our work, we
have used different values for these parameters and have plotted various diagrams for
the effective EoS, the deceleration parameter and the statefinder parameters which all
are calculated from the modified normalized Hubble parameter for each model. Note that,
we have used the present value for the baryonic density parameter which is about five
percent of the total matter density parameter. In each diagram we have indicated the
present values of the related quantities in order to compare the predictions of the
underlying model with the corresponding observational values. Particularly, we have
plotted the curves of quantities which hold in the GR background (i.e. when the
constants $\mathbf{m}$ and $\mathbf{n}$'s do vanish) in order to compare them with
the $f(R,T)$ corresponding plots and have investigated the possible deviations. We
mentioned that positive values for parameters $\mathbf{m}$ and $\mathbf{n}$'s
lead to some divergences in diagrams and consequently to non-physical outcomes. As a result,
we have not followed our considerations for positive values of these parameters. The effective EoS
parameter for our models starts from a zero value in the early stages of the
evolution of the Universe and converges to $-1$, as is expected. Note that
the models of GCG/SCG in the GR background show a similar behavior~\cite{stfi10}
(as we also depicted this behavior in the diagrams). However, in $f(R,T)$ gravity
the present values can be improved as compared with the corresponding GR ones. As we
have shown, in the GR background, only for cases with $K\rightarrow1$ we have
$w_{0}^{\textrm{(eff)}}\simeq-1$. Nevertheless, in $f(R,T)$ gravity this value
will improve. For example, the values of the EoS parameter for $K=0.33$ is
$w_{0}^{\textrm{(eff)(GR)}}\simeq-0.314$ and the corresponding values in
the three models are; $w_{0}^{\textrm{(eff)(I)}}\simeq-0.773$ for $\mathbf{m}=0$
and $\mathbf{n^{(I)}}=-1$, $w_{0}^{\textrm{(eff)(II)}}\simeq-0.916$ for
$\mathbf{m}=-1$, $\mathbf{n^{(II)}}=-1$ and $\alpha=0.99$, and
$w_{0}^{\textrm{(eff)(III)}}\simeq-0.526$ for $\mathbf{m}=0$, $\mathbf{n^{(III)}}=-1$ and
an arbitrary constant $\alpha$. Comparing these values we get another important point. For
models I and III, the best values are obtained for $\mathbf{m}=0$, which means the baryonic
term is absent. It shows that in these models the pure SCG/GCG can drive the accelerated
expansion of the Universe in $f(R,T)$ gravity. The better values have obtained in model II,
when a mixture of the baryonic matter and the GCG in the high pressure regime are included.
Note that, the larger constant $K$ becomes, the more observationally accepted values the
EoS parameter gets. It means that models I and III can still be compatible provided that
constants $K$ and $\alpha$ are selected properly. However, model II can give admissible
values even for smaller values of $K$. There is a special case; the values of
$w_{0}^{\textrm{(eff)}}$ and $q_{0}$ do not remain constant for $K=0$ when
$\mathbf{m}=-1$ in models I and II and arbitrary values of $\mathbf{m}$ for model III, (in
this case we must have $w_{0}^{\textrm{(eff)}}=0$ and $q_{0}=0.5$). In all models with
$K=0$, CG plays the role of CDM and therefore DE is absent in the history of the evolution
of the Universe. However, in $f(R,T)$ gravity, the mixture of DM and the baryonic matter
can lead to the accelerated expansion.

Also, we have investigated these three Chaplygin gas models in $(s, r)$ plane, where $s$ and $r$ are the
statefinder parameters. This tool would allow us to discriminate between different models of DE.
In the statefinder diagnosis, the difference between the predicted present values $s_{0}$
and $r_{0}$ of the models and the corresponding values for $\Lambda$CDM model (which is
called the distance to the $\Lambda$CDM model) is used as a criteria for discrimination of
DE models. Note that, we have $s^{(\Lambda CDM)}=0$ and $r^{(\Lambda CDM)}=1$, which
this fixed point is indicated by a green solid box in our diagrams. We have derived these
parameters for the three models and have plotted their evolutionary trajectories for
different values of $K$, $\mathbf{m}$, $\mathbf{n}$'s and $\alpha$. For the sake of clarity, we
have also drawn the GR corresponding results. In model I, the distance to the $\Lambda$CDM model
becomes shorter as compared with the same model with the GR background. For $\mathbf{m}=-1$ and
$\mathbf{n^{(I)}}=0$ both statefinder parameters slightly approach to the corresponding values
for $\Lambda$CDM model, and the minimum distance achieved when both $\mathbf{m}$ and
$\mathbf{n^{(I)}}$ get negative values. Moreover, we have shorter distances for
larger values of $K$, for all values of $\mathbf{m}$ and $\mathbf{n^{(I)}}$. The models
of class I, can be categorized in two different subclasses based on the initial values of the
statefinder parameter $s$; some models have trajectories starting from $s=-1/2$ which
correspond to $\mathbf{m}=0$. However, trajectories of some other models start from $s=+1/2$
with negative values of $\mathbf{m}$. Within the models which belong to class II, there are several
possibilities that can be considered as different scenarios of DE and can be classified
in three different categories based on their values of the statefinder parameters:
\begin{itemize}
  \item (i) There are Chaplygin gas models that are not effectively distinguishable from the $\Lambda \rm{CDM}$
  model. They have $|s^{(II)}-s^{(\Lambda \rm{CDM})}|,|r^{(II)}-r^{(\Lambda \rm{CDM})}|\lesssim10^{-3}$,
  corresponding to $\alpha\ll1$.
  \item (ii) Models that have long distance from the $\Lambda \rm{CDM}$ model, and they appear when
  $\mathbf{n^{(I)}}<0$ and $\alpha\rightarrow1$.
  \item (iii) The models that can be accounted for the third category and include
  large values of $\alpha$. Their distance are between those of categories (i) and (ii).
\end{itemize}
Note that, all cases with $K\rightarrow1$ have shorter distance than
other ones. In class III of models all trajectories start from $s=1/2$ in
the $(s,r)$ plane. The distance to the $\Lambda$CDM model is weakly affected
by variation of the value $\alpha$ when $\mathbf{n^{(III)}}<0$.
For $\mathbf{n^{(III)}}=0$, the trajectories are the same as those
models of class II and belong to categories (i) and (iii).
Finally, we have employed the chi-square test to examine models I, II and III
using recent observational data. Specifically, the Union 2 sample including 557
SNIa data and 28 Hubble parameter measurements have been used and the best fit
values of the model parameters by minimizing the so called chi-square function (it
can be accounted for as the function of all model parameters)
have been obtained. We found that in models I and II, the behavior of the modulus distance, the Hubble parameter
and the effective EoS parameter are consistent with the observations for those best fit parameters
which minimize the modulus distance chi-square. In this case we have
obtained $ \chi^{\textrm{2, (I)}}_{\textrm{$\mu$,min}}/d.o.f=1.0073$ and
$ \chi^{\textrm{2, (II)}}_{\textrm{$\mu$,min}}/d.o.f=1.0104$ which show admissible values. The best fit parameters gained from the Hubble
parameter chi-square test lead to a non-compatible value for the present effective EoS parameter and also
in this case the theoretical curve of modulus distance does not match the observational data in both models I and II. We compared
the values of statefinder parameters which are predicted by these three models by the observational data. It is shown that
the results from the observational constraints on the statefinder parameters are in favor of model II.
We came to conclusion that the results of performed studies in this work, for $f(R,T)$ gravity with GCG in
high pressure regimes can be reassuring enough to merit further investigations.
\section*{Acknowledgments}
The author acknowledges the university of Sistan and Bluchestan
for financial support, Grant No. 2016-952/2/348 and also is grateful
to Amir Hadi Ziaie for helpful discussions and correspondence.


\begin{thebibliography}{99}
\bibitem{supno1}   Riess, A.G., {\it et al.} ``Observational evidence from supernovae from an accelerating universe and a
                   cosmological constant", \textit{Astron. J.} \textbf{116} (1998), 1009.
\bibitem{supno2}   Perlmutter, S., {\it et al.} (The Supernova Cosmology Project), ``Measurements of $\Omega$ and $\Lambda$ from
                   $42$ high--redshift supernovae", \textit{Astrophys. J.} \textbf{517} (1999), 565.
\bibitem{supno3}   Riess, A.G., {\it et al.} ``BVRI curves for $22$ type Ia supernovae", \textit{Astron. J.} \textbf{117} (1999), 707.
\bibitem{WMAP1}    Spergel, D.N., {\it et al.} ``First year wilkinson microwave anisotropy probe (WMAP) observations: determination of cosmological
                   parameters", \textit{Astrophys. J. Suppl.} \textbf{148} (2003) ,175.
\bibitem{Teg1}     Tegmark, M., {\it et al.} ``Cosmological parameters from SDSS and WMAP", \textit{Phys. Rev. D} \textbf{69} (2004), 103501.
\bibitem{SDSS1}    Abazajian, K., {\it et al.} ``The second data release of the Sloan Digital Sky Survey", \textit{Astron. J.} \textbf{128} (2004), 502.
\bibitem{SDSS2}    Abazajian, K., {\it et al.} ``The third data release of the Sloan Digital Sky Survey ", \textit{Astron. J.} \textbf{129} (2005), 1755.
\bibitem{WMAP2}    Spergel, D.N., {\it et al.} ``Wilkinson microwave anisotropy probe (WMAP) three year results: implications for
                   cosmology", \textit{Astrophys. J. Suppl.} \textbf{170} (2007) ,377.
\bibitem{WMAP3}    Komatsu, E, {\it et al.} ``Five-Year wilkinson microwave anisotropy probe (WMAP) observations: cosmological interpretation",
                   \textit{Astrophys. J. Suppl.} \textbf{180} (2009) ,330.
\bibitem{WMAP4}    Komatsu, E, {\it et al.} ``Seven-Year wilkinson microwave anisotropy probe (WMAP) observations: cosmological interpretation",
                   \textit{Astrophys. J. Suppl.} \textbf{192} (2011) ,18.
\bibitem{WMAP5}    Hinshaw, G.F., {\it et al.} ``Nine-Year wilkinson microwave anisotropy probe (WMAP) observations: cosmological parameter results",
                   \textit{Astrophys. J. Suppl.} \textbf{208} (2013) ,19.
\bibitem{dmatt1}   Bertonea, G., Hooperb, D. \& Silk, J. ``Particle dark matter: evidence, candidates and constraints",
                   \textit{Phys. Rep.} \textbf{405} (2005), 279.
\bibitem{dmatt2}   Silk, J. ``Dark matter and galaxy formation", \textit{Ann. Phys. (Berlin)} \textbf{15} (2006), 75.
\bibitem{dmatt3}   Feng, J.L. ``Dark matter candidates from particle physics and methods of detection",
                   \textit{Annu. Rev. Astron. Astrophys.} \textbf{48} (2010), 495.
\bibitem{dmatt4}   Bergstr\"{o}m, L. ``Dark matter evidence, particle physics candidates and detection methods",
                   \textit{Ann. Phys. (Berlin)} \textbf{524} (2012), 479.
\bibitem{dmatt5}   Frenk, C.S. \& White, S.D.M. ``Dark matter and cosmic structure", \textit{Ann. Phys. (Berlin)} \textbf{524} (2012), 507.
\bibitem{dmatt6}   Destri, C., de Vega, H.J., \& Sanchez, N.G. ``Warm dark matter primordial spectra and the onset
                   of structure formation at redshift z", \textit{Phys. Rev. D} \textbf{88} (2013), 083512.
\bibitem{dmatt7}   Cheng, D., Chu, M.-C. \& Tang, J. ``Cosmological structure formation in decaying
                   dark matter models", \textit{J. Cosmol. Astropart. Phys.} \textbf{07} (2015), 009.
\bibitem{dmatt8}   Vogelsberger, M. {\it et al.} ``ETHOS - An effective theory of structure formation:
                   dark matter physics as a possible explanation of the small-scale CDM problems",
                   \textit{Mon. Not. R. Astron. Soc.} \textbf{460}, (2016), 1399.
\bibitem{dener1}   Peebles, P.J.E. ``The cosmological constant and dark energy", \textit{Rev. Mod. Phys.} \textbf{75} (2003), 559.
\bibitem{dener2}   Polarski, D. ``Dark energy: current issues", \textit{Ann. Phys. (Berlin)} \textbf{15} (2006), 342.
\bibitem{dener3}   Copeland, E.J., Sami, M. \& Tsujikawa, S. ``Dynamics of dark energy", \textit{Int. J. Mod. Phys. D}
                   \textbf{15} (2006), 1753.
\bibitem{dener4}   Durrer, R. \& Maartens, R. ``Dark energy and dark gravity: theory overview", \textit{Gen. Rel. Grav.}
                   \textbf{40} (2008), 301.
\bibitem{dener5}   Rosales, J.J. \& Tkach, V.I. ``Supersymmetric cosmological FRW model and dark energy",
                   \textit{Phys. Rev. D} \textbf{82} (2010), 107502.
\bibitem{dener6}   He, J.-H., Wang, B. \& Abdalla, E. ``Testing the interaction between dark energy and dark matter
                   via latest observations", \textit{Phys. Rev. D} \textbf{83} (2011), 063515.
\bibitem{dener7}   Xia, J.-Q. ``New limits on coupled dark energy from Planck", \textit{J. Cosmol. Astropart. Phys.} \textbf{11} (2013), 022.
\bibitem{dener8}   Yang, W. \& Xu, L. ``Coupled dark energy with perturbed Hubble expansion rate",
                   \textit{Phys. Rev. D} \textbf{90} (2014), 083532.
\bibitem{dener9}   Linde, A., Roest, D. \& Scalisi, M. ``Inflation and Dark Energy with a Single Superfield",
                   \textit{J. Cosmol. Astropart. Phys.} \textbf{03} (2015), 017.
\bibitem{dener10}  Odderskov, I., Baldi, M. \& Amendo, L. ``The effect of interacting dark energy on local measurements of
                   the Hubble constant", \textit{J. Cosmol. Astropart. Phys.} \textbf{05} (2016), 035.
\bibitem{dener11}  Wang, B., Abdalla, E., Atrio-Barandela, F., Pavon, D. ``Dark matter and dark energy interactions:
                   Theoretical challenges, cosmological implications and observational signatures", arXiv:1603.08299 [astro-ph].
\bibitem{dener12}  Dodelson, S. {\it et al.} ``Cosmic Visions Dark Energy: Science", arXiv:1604.07626 [astro-ph].
\bibitem{Planck1}  Ade, P.A.R. {\it et al.} ``Planck 2013 results. XVI. cosmological parameters", \textit{A}\&\textit{A}
                   \textbf{571} (2013), A16.
\bibitem{Planck2}  Ade, P.A.R. {\it et al.} ``Planck 2015 results. XIII. cosmological parameters", \textit{astro-ph/1502.01589}.
\bibitem{LCDM}     Ostriker, J.P. \& Steinhardt, P.J. ``Cosmic concordance'', \textit{astro-ph/9505066}.
\bibitem{Cos.pro1} Weinberg, S. ``The cosmological constant problem'', \textit{Rev. Mod. Phys.} \textbf{61} (1989), 1.
\bibitem{Cos.pro2} Nobbenhuis, S. ``The cosmological constant problem, an inspiration for new physics'', \textit{gr-qc/0609011}.
\bibitem{Cos.pro3} Padmanabhan, H. \& Padmanabhan, T. ``CosMIn: The solution to the cosmological constant problem'',
                   \textit{Int. J. Mod. Phys. D} \textbf{22} (2013), 1342001.
\bibitem{Cos.pro4} Bernard, D. \& LeClair, A. ``Scrutinizing the cosmological constant problem and a possible resolution'',
                   \textit{Phys. Rev. D} \textbf{87} (2013), 063010.
\bibitem{Cos.pro5} Narimani, A., Afshordi, N. \& Scott, D. ``How does pressure gravitate? Cosmological constant
                   problem confronts observational cosmology", \textit{J. Cosmol. Astropart. Phys.} \textbf{08} (2014), 049.
\bibitem{Cos.pro6} Moffat, J.W. ``Quantum gravity and the cosmological constant problem", \textit{gr-qc/1407.2086}.
\bibitem{Cos.pro7} Padilla, A. ``Lectures on the cosmological constant problem", \textit{hep-th/1502.05296}.
\bibitem{Cos.pro8} Linder, E.V. `` Quintessence's last stand?", \textit{Phys. Rev. D} \textbf{91} (2015), 063006.
\bibitem{quint1}   Chiba, T., Okabe, T. \& Yamaguchi, M. ``Kinetically driven quintessence'', \textit{Phys. Rev. D} \textbf{62} (2000), 023511.
\bibitem{quint2}   Armendariz-Picon, C., Mukhanov, V. \& Steinhardt, P.J. ``Dynamical solution to the problem of a small cosmological
                   constant and late--time cosmic acceleration'', \textit{Phys. Rev. Lett.} \textbf{85} (2000), 4438.
\bibitem{quint3}   Saridakis, E.N. \& Ward, J. ``Quintessence and phantom dark energy from ghost D-branes",
                   \textit{Phys. Rev. D} \textbf{80} (2009), 083003.
\bibitem{quint4}   Sheykhi, A. \& Bagheri, A. ``Quintessence ghost dark energy model", \textit{Europhys. Lett.} \textbf{95} (2011 ), 39001.
\bibitem{quint5}   B\"{o}hmer, C.G., Tamanini, N. \& Wright, M. ``Interacting quintessence from a variational
                   approach. II. Derivative couplings", \textit{Phys. Rev. D} \textbf{91} (2015), 123003.
\bibitem{quint6}   Li, D. \& Scherrer, R.J. ``Classifying the behavior of noncanonical
                   quintessence", \textit{Phys. Rev. D} \textbf{92} (2015), 083509.
\bibitem{CG1}      Kamenshchik, A., Moschella, U., \& Pasquier, V. ``An alternative to quintessence",
                   \textit{Phys. Lett. B} \textbf{511} (2001 ), 265.
\bibitem{CG2}      Fabris, J.C., Goncalves, S.V.B., \& de Souza, P.E. ``Density perturbations in
                   a universe dominated by the Chaplygin gas", \textit{Gen. Rel. Grav.} \textbf{34} (2002), 53.
\bibitem{CG3}      Bento, M.C.,  Bertolami, O., \& Sen, A.A. ``Generalized Chaplygin gas, accelerated expansion
                   and dark energy-matter unification", \textit{Phys. Rev. D} \textbf{66} (2002), 043507.
\bibitem{CG4}      Avelino, P.P., Be�a, L.M.G., de Carvalho, J.P.M., Martins, C.J.A.P. \& Pinto, P. ``Alternatives to quintessence model
                   building", \textit{Phys. Rev. D} \textbf{67} (2003), 023511.
\bibitem{CGprob1}  Bean, R., \& Dore, O. ``Are Chaplygin gases serious contenders for the dark energy?",
                   \textit{Phys. Rev. D} \textbf{68} (2003), 023515.
\bibitem{CGprob2}  Sandvik, S.B., Tegmark, M., Zaldarriaga, M., \& Waga, I. ``The end of unified dark matter?"
                   \textit{Phys. Rev. D} \textbf{69} (2004), 123524.
\bibitem{GCG1}     Dindam, B.R., Kumar, S., \& Sen, A.A. ``Inflationary generalized Chaplygin gas and dark energy
                   in the light of the Planck and BICEP2 experiments", \textit{Phys. Rev. D} \textbf{90} (2014), 083515.
\bibitem{GCG2}     Ebadi, H., \& Moradpour, H. ``Thermodynamical description of modified generalized Chaplygin gas
                   model of dark energy", \textit{Int. J. Theor. Phys.} \textbf{55} (2016), 1612.
\bibitem{GCG3}     Fabris, J.C., Goncalves, S.V.B., Velten, H.E.S., \& Zimdahl, W. ``Matter power spectrum
                   for the generalized Chaplygin gas model: The newtonian approach", \textit{Phys. Rev. D} \textbf{78} (2008), 103523.
\bibitem{GCG4}     Eiroa, E.F. ``Thin-shell wormholes with a generalized Chaplygin gas", \textit{Phys. Rev. D} \textbf{80} (2009), 044033.
\bibitem{GCG5}     Bejarano, C., \& Eiroa, E.F. ``Dilaton thin-shell wormholes supported by a generalized
                   Chaplygin gas", \textit{Phys. Rev. D} \textbf{84} (2011), 064043.
\bibitem{GCG6}     Kuhfittig, P.K.F. ``Wormholes admitting conformal Killing vectors and supported by generalized
                   Chaplygin gas", \textit{Eur. Phys. J. C} {\bf 75} (2015), 357.
\bibitem{fR1}      Nojiri, S. \& Odintsov, S.D. ``Introduction to modified gravity and gravitational alternative for
                   dark energy'', \textit{Int. J. Geom. Meth. Mod. Phys.} \textbf{04} (2007), 115.
\bibitem{fR2}      De Felice, A. \& Tsujikawa, S. ``$f(R)$ theories", \textit{Living Rev. Rel.} \textbf{13} (2010), 3.
\bibitem{fR3}      Sotiriou, T.P. \& Faraoni, V. ``$f(R)$ theories of gravity", \textit{Rev. Mod. Phys.} \textbf{82} (2010), 451.
\bibitem{fR4}      Ziaie, A.H., Atazadeh, K., \& Rasouli, S.M.M. ``Naked Singularity Formation In $f(R)$ Gravity",
                   \textit{Gen. Rel. Grav.} \textbf{43} (2011), 2943.
\bibitem{fR5}      Jennings, E., Baugh, C.M., Li, B., Zhao, G.-B. \& Koyama, K. ``Redshift-space distortions
                   in $f(R)$ gravity", \textit{Mon. Not. R. Astron. Soc.} \textbf{425}, (2012), 2128.
\bibitem{fR6}      Carloni, S., Goswami, R. \& Dunsby, P.K.S. ``A new approach to reconstruction methods in
                   $f(R)$ gravity", \textit{Class. Quantum Grav.} \textbf{29}, (2012), 135012.
\bibitem{fR7}      Aviles, A., Bravetti, A., Capozziello, S. \& Luongo, O. ``Updated constraints on $f(R)$ gravity
                   from cosmography", \textit{Phys. Rev. D} \textbf{87} (2013), 044012.
\bibitem{fR8}      Makarenko, Andrey N., Odintsov, S. \& Olmo, Gonzalo J. ``Born-Infeld $f(R)$ gravity", \textit{Phys. Rev. D} \textbf{90} (2014), 024066.
\bibitem{fR9}      Cataneo, M., {\it et al.} ``New constraints on $f(R)$ gravity from clusters of
                   galaxies", \textit{Phys. Rev. D} \textbf{92} (2015), 044009.
\bibitem{fR10}     Battye, R.A., Bolliet, B. \& Pearson, J.A. ``$f(R)$ gravity as a dark energy fluid", \textit{Phys. Rev. D} \textbf{93} (2016), 044026.
\bibitem{frL1}     Bertolami, O., B$\ddot{o}$hmer, C.G., Harko, T. \& Lobo, F.S.N. ``Extra force in $f(R)$ modified theories
                   of gravity'', \textit{Phys. Rev. D} \textbf{75} (2007), 104016.
\bibitem{frL2}     Bertolami, O. \& P$\acute{a}$ramos, J. ``Do $f(R)$ theories matter?'', \textit{Phys. Rev. D} \textbf{77} (2008), 084018.
\bibitem{frL3}     Bertolami, O., Lobo, F.S.N., \& P$\acute{a}$ramos, J. ``Nonminimal coupling of perfect fluids
                   to curvature'', \textit{Phys. Rev. D} \textbf{78} (2008), 064036.
\bibitem{frL4}     Harko, T. ``Modified gravity with arbitrary coupling between matter and geometry'', \textit{Phys. Lett. B}
                   \textbf{669} (2008), 376.
\bibitem{frL5}     Nesseris, S. ``Matter density perturbations in modified gravity models with arbitrary
                   coupling between matter and geometry'', \textit{Phys. Rev. D} \textbf{79} (2009), 044015.
\bibitem{frL6}     Harko, T. \& Lobo, F.S.N. ``$f(R,L_{m})$ gravity'', \textit{Eur. Phys. J. C} \textbf{70} (2010), 373.
\bibitem{fRT1}     Harko, T, Lobo, Francisco S.N., Nojiri, S. \& Odintsov, Sergei D. ``$f(R,T)$
                   gravity", \textit{Phys. Rev. D} \textbf{84} (2011), 024020.
\bibitem{fRT2}     Houndjo, M.J.S., Alvarenga, F.G., Rodrigues, M.E., Jardim, D.F. \& Myrzakulov, R. ``Thermodynamics in little
                   rip cosmology in the framework of a type of $f(R,T)$ gravity", \textit{gr-qc/1207.1646}.
\bibitem{fRT3}     Sharif, M. \& Zubair, M. ``Thermodynamics in $f(R,T)$ theory of gravity", \textit{J. Cosmol. Astropart.
                   Phys.} \textbf{03} (2012), 028.
\bibitem{fRT4}     Jamil, M., Momeni, D. \& Ratbay, M. ``Violation of the first law of thermodynamics in $f(R,T)$ gravity",
                   \textit{Chin. Phys. Lett.} \textbf{29} (2012), 109801.
\bibitem{fRT5}     Sharif, M. \& Zubair, M. ``Thermodynamic behavior of particular $f(R,T)$-gravity models", \textit{J. Exp. Theor. Phys.}
                   \textbf{117} (2013), 248.
\bibitem{fRT6}     Alvarenga, F.G., Houndjo, M.J.S., Monwanou, A.V. \& Chabi Orou, J.B. ``Testing some $f(R,T)$ gravity models from energy
                   conditions", \textit{J. Mod. Phys.} {\bf 04} (2013), 130.
\bibitem{fRT7}     Shabani, H. \& Farhoudi, M. ``$f(R,T)$ cosmological models in phase-space", \textit{Phys. Rev. D} \textbf{88}
                   (2013), 044048.
\bibitem{fRT8}     Sharif, M. \& Zubair, M. ``Energy conditions in $f(R,T,R_{\mu \nu}T^{\mu \nu})$ gravity", \textit{J. High Energy Phys.}
                   {\bf 12} (2013), 079.
\bibitem{fRT9}     Kiani, F. \& Nozari, K. ``Energy conditions in $F(T,\Theta)$ gravity and compatibility with a stable de
                   Sitter solution", \textit{Phys. Lett. B} {\bf 728} (2014), 554
\bibitem{fRT10}    Shabani, H. \& Farhoudi, M. ``cosmological and solar system consequences of $f(R,T)$ gravity models",
                   \textit{Phys. Rev. D} \textbf{90} (2014), 044031.
\bibitem{fRT11}    Noureen, I. \& Zubair, M. ``Dynamical instability and expansion-free condition in $f(R,T)$
                   gravity", \textit{Eur. Phys. J. C} {\bf 75} (2015), 62.
\bibitem{fRT12}    Azizi, T. \& Yaraie, E. ``G\"{o}del-type universes in Palatini $f(R)$ gravity with a non-minimal
                   curvature-matter coupling", \textit{Int. J. Theor. Phys.} {\bf 55} (2016), 176.
\bibitem{fRT13}    Jos� Barrientos, O. \& Rubilar, Guillermo F. ``Surface curvature singularities of polytropic spheres in Palatini
                   $f(R,T)$ gravity", \textit{Phys. Rev. D} \textbf{93} (2016), 024021.
\bibitem{fRT14}    Alves, M.E.S., Moraes, P.H.R.S., de Araujo, J.C.N. \& Malheiro, M. ``Gravitational waves in
                   the $f(R,T)$ theory of gravity", \textit{gr-qc/1604.03874}.
\bibitem{fRT15}    Shabani, H. \& Ziaie, A.H.``Stability of the Einstein static universe in $f(R,T)$ gravity`", \textit{gr-qc/1606.07959}.
\bibitem{riess}    Riess, A.G., {\it et al.} ``Type Ia supernova discoveries at $z>1$ from the Hubble Space Telescope:
                   evidence for past deceleration and constraints on dark energy evolution", \textit{Astrphys. J.} \textbf{607} (2004), 665.
\bibitem{astier}   Astier, P., {\it et al.} ``The Supernova Legacy Survey: measurement of
                   $\Omega_{M}$, $\Omega_{\Lambda}$ and $w$ from the first year data set", \textit{A\&A} \textbf{447} (2006), 31.
\bibitem{rapetti}  Rapetti, D., {\it et al.} ``A kinematical approach to dark energy studies",
                   \textit{Mon. Not. R. Astron. Soc.} \textbf{375}, (2007), 1510.
\bibitem{rul1}    Wu, P. \& Yu, H. ``Constraints on the unified dark energy�dark matter model from latest
                  observationaldata", \textit{J. Cosmol. Astropart. Phys.}, \textbf{03} (2007), 015
\bibitem{rul2}    Bento, M.C.,  Bertolami, O., \& Sen, A.A. ``Generalized Chaplygin gas and cosmic
                  microwave background radiation constraints", \textit{Phys. Rev. D} \textbf{67} (2003), 063003.
\bibitem{rul3}    Zhu, Z.-H. ``Generalized Chaplygin gas as a unified scenario of dark
                  matter/energy: Observational constraints", \textit{A\&A} \textbf{423} (2004), 421.
\bibitem{rul4}    Wu, P., \& Yu, H. ``Generalized Chaplygin gas model: Constraints from Hubble parameter
                  versus redshift data", \textit{Phys. Lett. B}, \textbf{644} (2007), 16.
\bibitem{rul5}    Wu, P., \& Yu, H. ``Interacting generalized Chaplygin gas", \textit{Class. Quantum Grav.}, \textbf{24} (2007), 4661.
\bibitem{rul6}    Amendola, L., Waga I. \& Finelli, F., ``Observational constraints on silent
                  quartessence", \textit{J. Cosmol. Astropart. Phys.}, \textbf{11} (2005), 009.
\bibitem{elmardi}  Elmardi, M., Abebe, A. \& Tekola, A. ``Chaplygin-gas Solutions of $f(R)$ Gravity", \textit{gr-qc/1603.05535}.
\bibitem{borowiec} Borowiec, A., Stachowski, A., Szydlowski, M. \& Wojnar, A.``Inflationary cosmology
                   with Chaplygin gas in Palatini formalism", \textit{J. Cosmol. Astropart. Phys.}, \textbf{01} (2016), 040.
\bibitem{sandvik}  Sandvik, H{\aa}vard B., Tegmark, M., Zaldarriaga, M. \& Waga, I. ``The end of unified
                   dark matter?", \textit{Phys. Rev. D} \textbf{69} (2004), 123524.
\bibitem{popolo}   Popolo, A. Del, {\it et al.} ``Shear and rotation in Chaplygin cosmology", \textit{Phys. Rev. D} \textbf{87} (2013), 043527.
\bibitem{stfi1}    Sahni, V., Saini, T.D., Starobinsky, A.A., \& Alam, U. ``Statefinder--a new geometrical diagnostic of dark energy",
                   \textit{JETP Lett.} {\bf 77} (2003), 201.
\bibitem{stfi2}    Alam, U., Sahni, V., Saini, T.D. \& Starobinsky, A.A. ``Exploring the expanding Universe and dark energy using the
                   statefinder diagnostic", \textit{Mon. Not. R. Astron. Soc.} \textbf{344}, (2003), 1057.
\bibitem{stfi3}    Zimdahl, W. \& Pavon, D. ``Letter: statefinder parameters for interacting dark energy",
                   \textit{Gen. Rel. Grav.} \textbf{36} (2004), 1483.
\bibitem{stfi4}    Zhang, X. ``Dark energy constraints from the cosmic age and supernova", \textit{Phys. Lett. B}, \textbf{611} (2005), 1.
\bibitem{stfi5}    Zhang, X. ``Statefinder diagnostic for holographic dark energy model",
                   \textit{Int. J. Mod. Phys. D}, \textbf{14} (2005), 1597.
\bibitem{stfi6}    Zhang, J., Zhang, X. \& Liu, H. ``Statefinder diagnosis for the interacting model of holographic dark energy",
                   \textit{Phys. Lett. B}, \textbf{659} (2008), 26
\bibitem{stfi7}    Setare, M.R., Zhang, J. \& Zhang, X. ``Statefinder diagnosis in a non-flat universe and the holographic
                   model of dark energy", \textit{J. Cosmol. Astropart. Phys.}, \textbf{0703} (2007), 007.
\bibitem{stfi8}   Chang, B.R., Liu, H.Y., Xu, L.X., Zhang, C.W. \& Ping, Y.L. ``Statefinder parameters for interacting
                  phantom energy with dark matter", \textit{J. Cosmol. Astropart. Phys.}, \textbf{0701} (2007), 016.
\bibitem{stfi9}   Shao, Y. \& Gui, Y. ``Statefinder parameters for tachyon dark energy model", \textit{Mod. Phys. Lett. A},
                  \textbf{23} (2008), 65.
\bibitem{stfi10}  Malekjani, M., Khodam-Mohammadi, A. \& Nazari-Pooya, N. ``Generalized Chaplygin gas model: cosmological
                  consequences and statefinder diagnosis", \textit{Astrophys. Space Sci.}, \textbf{334} (2011), 193.
\bibitem{stfi11}  Zhang, L., Cui, J., Zhang, J., Zhang, X. ``Interacting model of new agegraphic dark energy: cosmological
                  evolution and statefinder diagnostic", \textit{Int. J. Mod. Phys. D}, \textbf{19} (2010), 21.
\bibitem{stfi12}  Khodam-Mohammadi, A. \& Malekjani, M. ``Cosmic behavior, statefinder diagnostic and $w-w'$ analysis
                  for interacting new agegraphic dark energy model in non-flat universe", \textit{Astrophys. Space Sci.},
                  \textbf{331} (2010), 265.
\bibitem{haghani}  Haghani, Z., Harko, T., Lobo, F.S.N., Sepangi, H.R. \& Shahidi, S. ``Further matters in space--time geometry:
                   $f(R, T, R_{\mu\nu} T^{\mu\nu})$ gravity'', \textit{Phys. Rev. D} {\bf 88} (2013), 044023.
\bibitem{Sharif}  Sharif, M. \& Zubair, M. ``Cosmological reconstruction and stability in $f(R,T)$
                  gravity", \textit{Gen. Rel. Grav.} \textbf{46} (2014), 1723.
\bibitem{fR}      Capozziello, S. \& Faraoni, V., {\it Beyond Einstein Gravity}, (Springer, Netherlands, 2011).
\bibitem{LamdaT-1} Pradhan, A., Ahmed, N. \& Saha, B. ``Reconstruction of modified $f(R,T)$ with $\Lambda(T)$
                   gravity in general class of Bianchi cosmological models", \textit{Can. J. Phys.}, \textbf{93} (2015), 654.
\bibitem{LamdaT-2} Saho, P.K. \& Sivakumar, M. ``LRS Bianchi type-I cosmological model in $f(R,T)$ theory of gravity with $\Lambda(T)$",
                   \textit{Astrophys. Space Sci.}, \textbf{357} (2015), 60.
\bibitem{LamdaT-3} Poplawski, Nikodem J. ``A Lagrangian description of interacting dark energy", \textit{gr-qc/0608031}.
\bibitem{bertolami} Bertolami, O., Gil Pedro, F. \& Le Delliou, M. ``Dark energy�dark matter interaction and putative violation
                    of the equivalence principle from the Abell cluster A586", \textit{Phys. Lett. B} \textbf{654} (2007), 165.
\bibitem{union}    Amanullah, R., {\it et al.} ``Spectra and Hubble space telescope light curves of six type Ia
                   supernovae at 0.511< z< 1.12 and the Union2 compilation", \textit{Astrophys. J.} \textbf{716} (2010), 712.
\bibitem{farooq}   Farooq, O., {\it et al.} ``Hubble parameter measurement constraints on the redshift of the deceleration-acceleration
                   transition, dynamical dark energy, and space curvature", \textit{astro-ph/1607.03537}.
\bibitem{H1}      Simon, J., Verde, L., \& Jimenez, R. ``Constraints on the redshift dependence of the dark energy
                  potential", \textit{Phys. Rev. D} \textbf{71} (2005), 123001.
\bibitem{H2}      Stern, D., {\it et al.} ``Cosmic chronometers: Constraining the equation of
                  state of dark energy. II. a spectroscopic catalog of red galaxies in galaxy
                  clusters", \textit{Astrophys. J.} \textbf{188} (2010), 280.
\bibitem{H3}      Moresco, M., {\it et al.} ``Improved constraints on the expansion rate of the
                  Universe up to z~1.1 from the spectroscopic evolution of cosmic chronometers",
                  \textit{J. Cosmol. Astropart. Phys.}, \textbf{08} (2012), 006.
\bibitem{H4}      Blake, C., {\it et al.} ``The WiggleZ Dark Energy Survey: Joint measurements of the
                  expansion and growth history at $z < 1$", \textit{Mon. Not. R. Astron. Soc.} \textbf{425}, (2012), 405.
\bibitem{H5}      Font-Ribera, A., {\it et al.} ``Quasar-Lyman $\alpha$ forest cross-correlation from
                  BOSS DR11: Baryon acoustic oscillations", \textit{J. Cosmol. Astropart. Phys.}, \textbf{05} (2014), 027.
\bibitem{H6}      Delubac, T., {\it et al.} ``Baryon acoustic oscillations in the Ly$\alpha$ forest of BOSS DR11 quasars"
                  , \textit{A\&A} \textbf{574} (2015), A59.
\bibitem{H7}      Moresco, M., {\it et al.} ``A $6\%$ measurement of the Hubble parameter at $z\sim0.45$: direct
                  evidence of the epoch of cosmic re-acceleration", \textit{J. Cosmol. Astropart. Phys.}, \textbf{05} (2016), 014
\bibitem{H8}      Cuesta, J.A., {\it et al.} ``The clustering of galaxies in the SDSS-III Baryon Oscillation Spectroscopic
                  Survey: Baryon acoustic oscillations in the correlation function of LOWZ
                  and CMASS galaxies in Data Release 12", \textit{Mon. Not. R. Astron. Soc.} \textbf{457}, (2016), 1770.
\bibitem{pack1}   https://github.com/CosmologyTaskForce/CoChiSquare/tree/alpha; DOI:10.5281/zenodo.13197, (2014).
\bibitem{Math}    http://www.wolfram.com/
\bibitem{pyth}    https://www.python.org/psf/
\bibitem{pack2}   Perivolaropoulos, L. \& Nesseris, S., http://leandros.physics.uoi.gr/cosmofit.htm.
\end{thebibliography}
\end{document}